\documentclass[12pt]{article}
\usepackage{graphics,graphicx,epsfig}
\usepackage{amssymb,amsmath,amsfonts}
\usepackage{ifthen}

\def\Tr{\mbox{Tr}\,}

\def\hbar{\hspace{0pt}\raisebox{1pt}{$-$} \hspace{-7pt} h}

\def\5{\overline 5}

\newcommand{\be}{\begin{equation}}
\newcommand{\ee}{\end{equation}}
\newcommand{\bea}{\begin{eqnarray}}
\newcommand{\eea}{\end{eqnarray}}

\newcommand{\half}{\frac{1}{2}}
\newcommand{\inv}{{-1}}
\newcommand{\munu}{{\mu\nu}}

\newcommand{\lsim}{\,\raise.3ex\hbox{$<$\kern-.75em\lower1ex\hbox{$\sim$}}\,}
\newcommand{\gsim}{\,\raise.3ex\hbox{$>$\kern-.75em\lower1ex\hbox{$\sim$}}\,}

\newcommand{\GG}{\mathcal{G}}
\newcommand{\HH}{\mathcal{H}}

\newcommand{\LL}{\mathcal{L}}
\newcommand{\NN}{\mathcal{N}}

\newcommand{\OO}{\mathcal{O}}

\newcommand{\TeV}{\text{ TeV }}

\newcommand{\MeV}{\text{ MeV}}
\newcommand{\eff}{{\text{eff}}}

\begin{document}
\begin{titlepage}
\renewcommand{\thefootnote}{\fnsymbol{footnote}}
\begin{flushright}
SLAC--PUB--10457\\
May 2004\\
hep-ph/0405242\\
\end{flushright}
\vskip 2cm
\begin{center}
{\large\bf Composite Vector Mesons from QCD to the Little Higgs}
\vskip 1cm
{\normalsize
Maurizio Piai$^{1}$\footnote{The work of MP is supported in part by the U.S. Department of Energy under contract number DE-FG02-92ER-4074.}, Aaron Pierce$^{2,3}$\footnote{The work of AP is supported by the U.S. Department of Energy 
under contract number DE-AC03-76SF00515.}, Jay G. Wacker$^{3}$\footnote{JGW is supported by National Science Foundation Grant PHY-9870115 and by the Stanford Institute for Theoretical Physics.}\\
\vskip 0.5cm
1. Department of Physics\\
   Yale University\\
   New Haven, CT 06520\\
\vskip .1in
2. Theory Group \\
   Stanford Linear Accelerator Center\\
   Menlo Park, CA 94025\\
\vskip .1in
3. Institute for Theoretical Physics\\
   Stanford University\\
   Stanford, CA 94305\\
\vskip .1in
}
\end{center}
\vskip .5cm

\begin{abstract}
We review how the $\rho$ meson can be modeled in an effective theory 
and discuss the implications of applying this approach to heavier 
spin-one resonances.   Georgi's vector limit is explored, and its 
relationship to locality in a deconstructed extra-dimension 
is discussed.  We then apply the formalism for $\rho$'s 
to strongly coupled theories of electroweak symmetry breaking, studying 
the lightest spin-one techni-$\rho$ resonances.  Understanding these 
new particles in Little Higgs models can shed light on previously 
incalculable, ultraviolet sensitive physics, including the mass of 
the Higgs boson.  
\end{abstract}
\vskip .5cm
\end{titlepage}

%
\section{Introduction}
\renewcommand{\thefootnote}{\arabic{footnote}}
\setcounter{footnote}{0}
\setcounter{equation}{0}
\renewcommand{\theequation}{\thesection.\arabic{equation}}

Little Higgs models have recently rekindled interest in theories where the 
Higgs is a pseudo-Nambu-Goldstone boson 
(PNGB)\cite{PGBHiggs,LH,littlest,sp6}.   
These models can naturally be ultraviolet-completed into theories of 
strong dynamics, although linear sigma model UV completions are possible.
If completed into a theory of strong dynamics, the PNGBs arise when the gauge 
dynamics at a scale $\Lambda \sim$ 10 TeV  breaks global chiral 
symmetries of the theory. These PNGBs are not the only states associated 
with the strong dynamics --  a host of resonances are expected near the scale 
of strong coupling.  
The energy at the Large Hadron Collider (LHC) will be 
insufficient to directly explore 
the constituents of the strong dynamics; therefore, the question of more 
immediate interest is how to describe the low-lying states of the strong 
dynamics. It is challenging to study the phenomenology of the
theory in this regime because the dynamics are strongly coupled.

The same situation exists in QCD: the chiral Lagrangian accurately
models the interactions of the pions at the lowest energies; perturbation 
theory is a powerful tool at high energies,  but it
is difficult to discuss the interactions of the QCD resonances near the strong
coupling scale, such as the $\rho$.  
Historically, a variety of techniques have been
employed to investigate these resonances, including current algebra, 
QCD sum rules, dispersion relations, and hidden local symmetry. 
In this paper, we first discuss the $\rho$ mesons of QCD; the lessons learned 
are used in the subsequent treatment of techni-$\rho$ mesons in Little Higgs 
theories.

We use hidden local symmetry because this technique  is valid at low 
energies and small
$Q^2$, precisely the regime in which we are interested. 
In the hidden local symmetry approach, 
one writes an effective Lagrangian including the $\rho$ mesons, 
analogous to the traditional chiral Lagrangian written for the pions
\cite{HiddenLocal}.  Gauge symmetries are useful for describing light vector 
mesons and can be used to constrain the interactions of the $\rho$ \cite{Georgi:1989xy}. 
The lightness of the $\rho$ (relative to $4 \pi f_{\pi} \equiv \Lambda$) 
is crucial to success in this program.  While the separation between
$m_{\rho}$ and $\Lambda$ is not incredibly large, it is enough to be predictive.  
These predictions are qualitatively correct and, perhaps surprisingly, 
quantitatively not far from the experimentally measured values.  

Another asset of the language of hidden local symmetry is that it clearly
illuminates the possibility of an enhanced symmetry for QCD, 
reflected in a particular value of the $\rho$-pion coupling.  
This symmetry point is essentially the ``vector limit'' discussed in \cite{Georgi:1989xy}, 
and reviewed in our Sec.~\ref{Sec:QCDVector}.  As we will discuss, 
QCD is not too far from realizing this enhanced symmetry.
Reference~\cite{deconstruction} showed that a theory comprised 
of gauge groups and bi-fundamental fields can be mapped directly on 
to a description of a physical extra dimension.  In particular, notions 
that are traditionally associated with a physical dimension, such as 
locality, can be preserved in the theory space description.   We find that 
the vector limit is closely related to locality in theory space. 

Including the lightest spin-one resonances in the effective theory
allows us to address questions of how high energy physics affects 
the PNGBs.  In particular, we can examine the 
unitarization of the PNGB scattering.  We also find
that the incorporating the techni-$\rho$'s  into the effective field
theory associated with electroweak symmetry breaking (EWSB) can shed light on previously 
incalculable quantities in Little Higgs models.

The organization of the paper is as follows.  In Sec. \ref{Sec:QCDrho}
we review the couplings of the $\rho$ in the QCD chiral Lagrangian.
This gives a well-understood example for the hidden local 
symmetry approach and motivates our  discussion of techni-$\rho$'s 
and their properties in
more general theories  of strong dynamics.  For our discussion of
$\rho$'s in QCD, we draw heavily  upon the paper \cite{Georgi:1989xy}.
In our discussion of the weak-scale strong dynamics, we will emphasize 
the potential utility of (techni-)$\rho$'s in regulating quadratic 
divergences.  To motivate this point, we again 
turn towards QCD. In Sec.~\ref{Sec:QCDVector}, we introduce 
the {\it vector limit} of QCD.  In this limit, the $\rho$'s regulate 
the quadratic divergence in the charged pion mass that
arises from QED loops \cite{Harada}.  
Having laid a foundation with our  discussion of the QCD $\rho$ meson,
we then turn towards a discussion of the  techni-$\rho$ and its couplings.
We explain how techni-$\rho$ mesons can
analogously soften quadratic divergences in Little Higgs models in
Sec.~\ref{Sec:TR}.  This allows us to calculate previously UV-sensitive quantities, including the Higgs boson mass in the Littlest Higgs model. We also 
note that the inclusion of techni-$\rho$'s 
can have important consequences for studying vacuum alignment: in the vector 
limit, the vacuum of the $SU(6)/Sp(6)$ theory is unstable.
In Sec.~\ref{Sec:unitarity}, we discuss how unitarity is
modified by incorporating light resonances.    
We argue that the low scale of unitarity violation
in theories with many PNGBs is likely a signal of light 
scalar resonances rather than vector modes.

\section{ The Hidden Locality of Vector Mesons In QCD}
\setcounter{equation}{0}
\renewcommand{\theequation}{\thesection.\arabic{equation}}
\label{Sec:QCDrho}

We first review the coupling of vector mesons
in two flavor QCD in the limit of vanishing quark masses\cite{Georgi:1989xy}. 
The chiral Lagrangian that describes the
coupling of the $\rho$ to the light pions will only 
depend on a handful of parameters.  Among these is a parameter that 
vanishes in Georgi's vector limit.  In this limit, the leading
cut-off sensitivity to $m_{\pi^\pm}^2 - m_{\pi^0}^2$ also 
vanishes, leaving a residual piece that is not sensitive to
cut-off uncertainties.  We spend this section exploring this example 
before moving to TeV scale physics in the next section.

In QCD with two flavors, $\psi_L^i$ and $\psi_R{}_j$ ($i,j$ = 1,2), there
is a global $SU(2)_L \times SU(2)_R$ flavor symmetry
that rotates two quarks amongst themselves:
\begin{eqnarray}
\psi_L \rightarrow g_L \psi_L, \hspace{0.5in} \psi_R \rightarrow g_R \psi_R .
\end{eqnarray}
After the QCD gauge coupling goes strong, this chiral symmetry is 
broken, and the quarks form a condensate
\begin{eqnarray}
\langle \psi^i_L \psi_R{}_j\rangle \sim 4 \pi f^3  \Sigma^i_j .
\end{eqnarray}
Here $\Sigma$ parameterizes the QCD pions, which are Goldstone bosons 
of the broken global symmetry  
\begin{eqnarray}
SU(2)_{L}\,\times\,SU(2)_{R}\,\longrightarrow\,SU(2)_{V} \, ,
\end{eqnarray}
where $SU(2)_V$ is the unbroken isospin symmetry.  Under this symmetry, 
the transformation of the Goldstone boson field is
\begin{equation}
\Sigma \rightarrow g_{L} \Sigma g_{R}^{\dagger}.
\end{equation}
The linearized fluctuations of $\Sigma$ and global symmetry
transformations are given by
\begin{eqnarray}
\Sigma\,\equiv\,e^{2 i\pi/f} \hspace{0.6in}
g_L \equiv e^{i \alpha_L}, \hspace{0.4in} g_R \equiv e^{i \alpha_R}.
\end{eqnarray}
The linearized $\pi$ field transforms under the global symmetries
as:
\begin{eqnarray}
\delta \pi = \frac{f}{2} ( \alpha_L - \alpha_R ) +\cdots .
\end{eqnarray}
The vector and axial--vector transformations are distinguished by the 
relationship between $\alpha_L$ and $\alpha_R$.
For the axial--vector transformation, 
we have $\alpha_L=-\alpha_R \equiv \alpha_{A}$, while for the vector 
transformation,
we have $\alpha_L=\alpha_R \equiv \alpha_{V}$.  
Then $\delta \pi = f \alpha_A$.
The quarks are also charged under a weakly gauged group, $U(1)_{EM}$.  
$U(1)_{EM}$ is contained within
$SU(2)_V$, and is gauged in  the $\tau^3$ direction.  The symmetry structure 
of the theory, both global and local, is illustrated in  Fig. 1.  
To leading order, the chiral Lagrangian describing the pions is:
\begin{eqnarray}
\label{Eq: QCD1}
\LL_\eff \,=\,- \frac{1}{4} F_\munu^2 + 
\frac{f^2}{4}\,\Tr|D_\mu\Sigma|^2 +\cdots ,
\end{eqnarray}
where the covariant derivative is given by
\begin{eqnarray}
D_\mu \Sigma = \partial_\mu \Sigma +i e A_\mu[\tau^3,\Sigma]
\hspace{0.3in} \text{ with } \Tr \tau^a \tau^b = \half \delta^{ab}.
\end{eqnarray}
$\Sigma$ transforms like $(2,\bar{2})$ under $SU(2)_L \times SU(2)_R$.  
Parity interchanges $SU(2)_L\leftrightarrow SU(2)_R$ and 
takes $\Sigma \rightarrow \Sigma^\dagger$.
Note that the $f$ in Eq.~\ref{Eq: QCD1}  can be identified with the pion 
decay constant, $f_\pi= 93$ MeV.

\subsection{Incorporating the $\rho$}
\label{Sec: QCD Rho}
\begin{figure}
\begin{center}
\epsfig{file=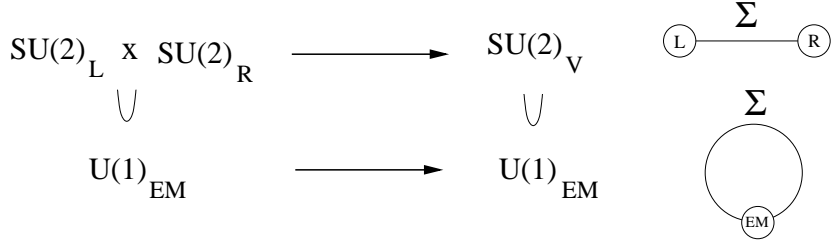, width=4.3in}
\caption{
\label{Fig: QCD1}
QCD with two flavors has a global $SU(2)_{L} \times SU(2)_{R}$ chiral 
symmetry, which is broken down to the diagonal $SU(2)_V$ (isospin) 
when the quarks condense.  As shown above, the $U(1)_{EM}$ gauge 
symmetry is contained within $SU(2)_V$.  We have also shown 
these symmetries in a ``sites and links'' theory space representation.  
The top diagram shows that the quark condensate transforms as a 
bi-fundamental under $SU(2)_L \times SU(2)_R$, 
and breaks this symmetry down to the diagonal.  The bottom diagram 
shows that the quark condensate is an adjoint of the gauged $ U(1)_{EM}$.
}
\end{center}
\end{figure}
Now we include the $\rho$ meson in our low energy theory.
The lightness of the $\rho$ mesons also motivates a
description utilizing  a gauge invariance, $SU(2)_\rho$.  The 
longitudinal  components of the $\rho$ are kept
explicit, and the gauge invariance can  be used to determine the
natural sizes of operators in the effective  Lagrangian.  Of course
this gauge symmetry has no physics in it, and  going to the unitary
gauge makes this clear.  A lucid discussion of this point was given
in \cite{Georgi:1989xy,Arkani-Hamed:2002sp}.   We emphasize that it is 
the lightness of the $\rho$ that constrains its properties-- 
we expect additional operators in the effective Lagrangian 
proportional to $m_{\rho}/\Lambda$. 
For example, there are higher derivative operators that can sum 
into a form-factor that reveals the composite nature of the $\rho$.
Upon including the $\rho$ in the chiral Lagrangian, the symmetry 
structure is enlarged to become:
\bea
SU(2)_{L}\,\times\,SU(2)_{R}\,\times\,SU(2)_{\rho}\,
\longrightarrow\,SU(2)_{V},
\eea
where the $SU(2)_\rho$ symmetry is strongly gauged.  This structure is 
displayed in Fig. 2.
\begin{figure}
\label{Fig: QCD2}
\begin{center}
\epsfig{file=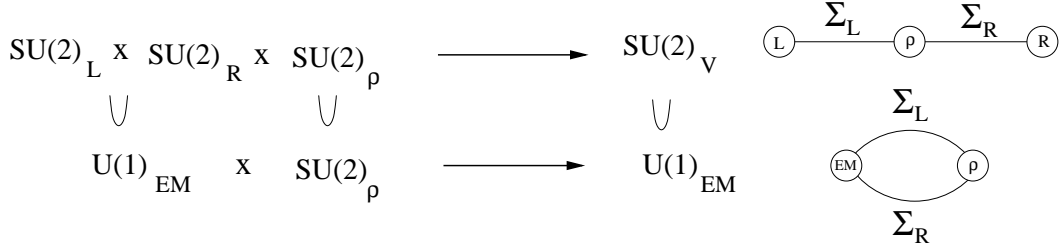, width=5.5in}
\caption{
The global (including $g_\rho \rightarrow 0$) and gauge symmetry 
structure of the QCD pions and a typical $\rho$ vector meson.  
This figure is the analogue of Fig. 1, now with 
the $\rho$'s included. The ``sites and links'' diagrams show  
the global and gauge symmetries under which the $\Sigma_{L}$ 
and $\Sigma_{R}$ fields transform as bi-fundamentals. 
}
\end{center}
\end{figure}

The effective theory, now incorporating both $\pi$'s and $\rho$'s, is
\begin{eqnarray}
\label{Eq: effLrhopi}
\nonumber
\LL_\eff\,&=&\,
-\frac{1}{4} F_\munu^2 - \half \Tr \rho_\munu^2 
- \tilde{\kappa} F_\munu(\Tr \tau^3 \Sigma_L \rho^\munu \Sigma_L^\dagger 
+\Tr \tau^3 \Sigma_R^\dagger \rho^\munu \Sigma_R)\\
&&
+\frac{f^2}{4}\,\left( \Tr|D^\mu\Sigma_L|^2
+\Tr|D^\mu \Sigma_R|^2
+\,\frac{\kappa}{2}\,\Tr|D^\mu(\Sigma_L\Sigma_R)|^2\right)\, +\cdots ,
\end{eqnarray}
where $\tilde{\kappa}$ sets the size of the $\rho-\gamma$ kinetic mixing.   
This effective action is the result of integrating out all heavier 
resonances.  We refer to 
the operators multiplied by $\kappa$ and $\tilde{\kappa}$ as 
``non-local'': in the theory-space description of Fig. 2, these 
operators involve more than one site, and so are indeed non-local in 
theory space.  For simplicity, we will set $\tilde{\kappa}$ to zero 
throughout; however, retaining 
it would be important if our goal were to try to match this theory to 
experimental QCD.
Parity acts on the theory by taking  $SU(2)_L \leftrightarrow SU(2)_R$,
leaving $SU(2)_\rho$ unchanged, and taking the Goldstone bosons from
$\Sigma_L\leftrightarrow \Sigma_R^\dagger .$
We have imposed parity on the Lagrangian in Eq. \ref{Eq: effLrhopi}.
The transformations under the $SU(2)_{L} \times SU(2)_{R} \times SU(2)_\rho$ 
symmetry are:
\begin{eqnarray}
\nonumber
\Sigma_{L} \equiv \exp(2 i \pi_L/f) 
\sim (2,\bar{2},1); &&  \Sigma_L \rightarrow  g_L\Sigma_L g_\rho^\dagger,\\
\nonumber
\Sigma_{R} \equiv \exp(2 i \pi_R/f) 
\sim\,(1,2,\bar{2}); &&  \Sigma_R \rightarrow  g_\rho\Sigma_R g_R^\dagger,\\
\Sigma_{L}\Sigma_{R}  \equiv \exp( 2 i (\pi_L+\pi_R)/f) 
\sim\,(2,1,\bar{2}); && \Sigma_L \Sigma_R 
\rightarrow g_L\Sigma_L\Sigma_R g_R^\dagger,
\end{eqnarray}
where $g_L$, $g_R$ and $g_\rho$ are independent $SU(2)$ transformations. 
The covariant derivatives that appear in Eq.~\ref{Eq: effLrhopi} are
\begin{eqnarray}
\nonumber
D_\mu  \Sigma_L &=& \partial_\mu \Sigma_L 
+ i e_0 A_\mu \tau^3\, \Sigma_L 
- i g_\rho \Sigma_L\, \vec{\tau}\cdot \vec{\rho}_\mu,\\
\nonumber
D_\mu  \Sigma_R &=& \partial_\mu\Sigma_R 
+ i g_\rho \vec{\tau}\cdot \vec{\rho}_\mu \,\Sigma_R 
-i e_0 \Sigma_R\, A_\mu \tau^3,\\
D_\mu  \Sigma_L \Sigma_R &=& \partial_\mu\Sigma_L \Sigma_R 
+ i e_0 A_\mu [\tau^3 ,\Sigma_L\Sigma_R].
\end{eqnarray}
Here, $A_\mu$ is the vector field that mixes with the $\rho^3$.  
Proceeding to the mass eigenbasis, we find a zero eigenvalue (the
photon)  and a massive eigenvalue (the physical $\rho$).  

It is possible to go to the unitary gauge where 
the new degrees of freedom become the longitudinal components of the
$\rho$.  This gauge is convenient because the physical couplings of
the $\rho$ are  manifest;  however, their natural sizes are
more difficult to  infer.   We can determine this gauge
by examining the Goldstone-$\rho$ mixing.
Ignoring the weakly gauged $U(1)_{EM}$, the mixing is given by
\begin{eqnarray}
\LL_{\rho \pi} = g_\rho f\,
\Tr \rho_\mu  \left(\partial^\mu\pi_L - \partial^\mu \pi_R\right) .
\end{eqnarray}
Thus, the physical pion, $\pi$, and the Goldstone
eaten by the $\rho$, denoted by $\xi$, are related to the gauge 
eigenstates by
\begin{eqnarray}
\pi = N_\pi^{-1}( \pi_L + \pi_R),
\hspace{0.5in}
\xi = N_\xi^{-1} (\pi_L - \pi_R),
\end{eqnarray}
where $N_\pi$ and $N_\xi$ are constants determined by the requirement that 
the fields be canonically normalized.  Unitary gauge is defined by $\xi =0$. 

The gauge and global transformations are
\begin{eqnarray}
g_L = e^{i \alpha_L},
\hspace{0.5in}
g_\rho = e^{i \alpha_\rho},
\hspace{0.5in}
g_R = e^{i \alpha_R},
\end{eqnarray}
and act upon the linearized fields as
\begin{eqnarray}
\delta \pi_L = \frac{f}{2} \Big( \alpha_L - \alpha_\rho\Big),
\hspace{0.5in}
\delta \pi_R = \frac{f}{2} \Big( \alpha_\rho - \alpha_R\Big).
\end{eqnarray}
The vector and axial vector transformations should preserve
unitary gauge\footnote{ 
This is a different definition of these global transformation 
than \cite{Son:2003et} used where both the vector and axial--vector 
transformations nominally took the theory out of unitary gauge.}.  
Therefore, we can parameterize the transformations
by the two global ones and the orthogonal one, which can be used to go to 
unitary gauge: 
\begin{eqnarray}
\rm{Vector:}\hspace{0.1in} \alpha_L  = \alpha_\rho = \alpha_R \equiv \alpha_V; 
\hspace{0.5in}
\rm{Axial:}  \hspace{0.1in}\alpha_L  = -\alpha_R \equiv  \alpha_A, \hspace{0.2in} \alpha_\rho =0.
\end{eqnarray}
The leading kinetic term  for the pions is given by
\begin{eqnarray}
\nonumber
\label{Eq: rho dpi2}
\LL_{(\partial\pi)^2} &=&
\Tr (\partial \pi_L)^2
+\Tr (\partial \pi_L)^2
+\kappa\Tr (\partial (\pi_L + \pi_R))^2;\\
&=& \half N_\pi^2(1 +\kappa)\Tr (\partial \pi)^2 + \half N_\xi^2\Tr (\partial \xi)^2
= \Tr (\partial \pi)^2 + \Tr (\partial \xi)^2,
\end{eqnarray}
meaning that the normalization constants are given by
\begin{eqnarray}
N_\pi^{-2} = \frac{1 +\kappa}{2},
\hspace{0.5in}
N_\xi^{-2} = \frac{1}{2}.
\end{eqnarray}
Under an axial transformation
\begin{eqnarray}
\delta \pi = N^{-1}_\pi f \alpha_A, \hspace{0.5in} \delta \xi =0.
\end{eqnarray}
Acting on Eq.~(\ref{Eq: rho dpi2}), we find:
\begin{eqnarray}
\delta \LL_{(\partial \pi)^2}
= 2 N^{-1}_\pi f \Tr \partial \pi \partial \alpha_A = 2 f_\pi \Tr \partial \pi \partial\alpha_A.
\end{eqnarray}
This allows us to identify
\begin{eqnarray}
f_\pi = \sqrt{\frac{1+\kappa}{2}} f .
\end{eqnarray}
The only \begin{it}a priori\end{it} 
constraint on $\kappa$ is that $\kappa > -1$, so that the physical $\pi$
has a positive kinetic term.  
At this point we can diagonalize the mass mixing with an orthogonal 
transformation
\begin{eqnarray}
A_{\text{Phys}} &=& \cos \theta \hat{A} + \sin \theta \rho^3,
\hspace{0.5in}
\rho^3_{\text{Phys}} = - \sin\theta \hat{A} + \cos \theta \rho^3.
\end{eqnarray}
where the angles and electro-magnetic gauge coupling are given by
\begin{eqnarray}
e^{-2}= e_0^{-2} + g_\rho^{-2},
\hspace{0.3in}
\sin \theta \equiv \frac{e}{g_\rho} .
\end{eqnarray}
So the couplings of the physical $\rho$ and photon to the 
electro-magnetic current, $j_{EM}$, are given by
\begin{eqnarray}
\label{Eq: Rho EM3}
\LL_{j\,\gamma} = e\, j_{\mu\;EM}\Big[ A^\mu_{\text{Phys}}
- \tan\theta \rho^{3\,\mu}_{\text{Phys}}\Big],
\end{eqnarray}
while the masses of the $\rho$ mesons are
\begin{eqnarray}
\label{Eq: Rhomasses}
m^2_{\rho^\pm} = \frac{ g^2_\rho f^2_\pi}{1+\kappa},
\hspace{0.2in}
m^2_{\rho^0} = \frac{ m^2_{\rho^\pm} }{\cos^2\theta}.
\hspace{0.3in}
\end{eqnarray} 
For large $g_\rho$, the difference between the masses of the
charged and neutral $\rho$ mesons can be expanded as
\begin{eqnarray}
\frac{m^2_{\rho^0}}{m^2_{\rho^\pm}} \approx 
\Big(1  + \frac{e^2}{g_\rho^2} \Big).
\end{eqnarray}
The experimental limits on the mass splittings are bounded to be
\begin{eqnarray}
m_{\rho^0}-m_{\rho^\pm}\lsim 1 \MeV \Rightarrow 1- \frac{m_{\rho^0}}{m_{\rho^\pm}} \le 3\times10^{-3}.\end{eqnarray}
For small $\tilde{\kappa}$, the mass splittings place a limit
on $g_\rho \simeq 4\pi/\sqrt{3}$.
Then, using $f_\pi \simeq 93 \MeV$ and 
$m_\rho \simeq 770\MeV$,  we find $\kappa \simeq \frac{1}{3}$ in QCD.
There are other determinations of $\kappa$ that give roughly the
same answer, e.g. using the KSFR relation for the $\rho\rightarrow \pi\pi$
decay width \cite{KSFR}.

For future reference we calculate the $\rho$'s  coupling to the isospin current of the $\pi$'s
\begin{eqnarray}
\label{Eq: rho Current Int}
\LL_{\text{int}} =  g j^a_\mu \rho_a^\pm \hspace{0.5in}
j^a_\mu = \Tr \tau^a [\partial_\mu \pi,\pi]
\hspace{0.3in} g = \frac{g_\rho}{1+\kappa}.
\end{eqnarray}

\subsection{Georgi's Vector Limit}

\label{Sec:QCDVector}

In this section, we discuss an enhanced symmetry of the strong dynamics 
known as the vector limit.  When this symmetry 
is exact, the $\rho$ meson acts to cut off the one loop gauge (QED) 
quadratic divergence of the charged pion mass.  
As we will discuss, a similar enhanced symmetry is possible in general 
theories of strong dynamics, including theories at the weak scale.

Starting from Eq.~\ref{Eq: effLrhopi}, one can compute the contribution 
of the photon to the mass of the charged pion with the Coleman-Weinberg potential
\cite{ColemanWeinberg}.  At one loop one finds: 
\begin{eqnarray}
\delta m^2_{\pi^\pm} = \frac{3e^2}{16\pi^2 (1 + \kappa)}\Big(
\frac{\kappa\Lambda^2}{\cos^2 \theta}  +  m^2_{\rho^0} 
\log\Lambda^2 + \cdots \Big).
\end{eqnarray}
The term ``$\cdots$'' includes a logarithmically divergent 
piece that is proportional to $\kappa \tan^{2} \theta$, which 
is numerically very small.
For the special case of $\kappa$=0, the one loop quadratic divergence is 
absent, and the counter-term from high energy physics is only necessary to
cancel two loop quadratic divergences.   The degree to which
the one-loop logarithmic divergence is larger than the two-loop 
quadratic divergence is the degree to which 
the $\delta m_{\pi^\pm}^2$ is calculable.  When $\kappa\ne 0$, a 
one loop quadratic divergence remains. 

This can be explained by studying the symmetry structure of the Lagrangian.  
For the QCD Lagrangian of Eq.~\ref{Eq: effLrhopi}, 
in the limit that $g_\rho$ and $e_{0}$ vanish, the global 
symmetry of the theory is as shown in Fig. 3.   Note 
if $\kappa=0$, this Lagrangian would allow independent transformations of the 
form $\Sigma_L \rightarrow \Sigma_L U_L$ and $\Sigma_R \rightarrow U_{R} \Sigma_R$. 
A non-zero $\kappa$ forces $U_L = U_R$.  
We refer to this limit of enhanced symmetry as the 
vector limit\footnote{
This is a slightly different definition of the vector limit taken in
\cite{Georgi:1989xy} where $\kappa$ \begin{it}and\end{it} $g_\rho$ 
were taken to vanish simultaneously.}. 

\begin{figure}
\begin{center}
\epsfig{file=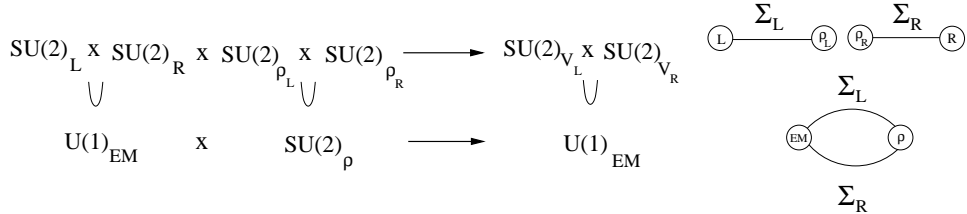, width=5in}
\caption{
\label{Fig: QCD3}
The global (including $g_\rho \rightarrow 0$) and gauge symmetry 
structure of the QCD pions 
and $\rho$ vector meson in the vector limit ($\kappa \rightarrow 0$).   
While the vector limit does not change the gauge symmetry structure of 
the theory, it does change the global symmetry. This 
translates into a relation between $m_\rho$, $g_\rho$ and $f_\pi$. 
}
\end{center}
\end{figure}

The global symmetry of the vector limit is
\begin{eqnarray}
SU(2)_L \times SU(2)_R \times SU(2)_{\rho_L} \times SU(2)_{\rho_R},  
\end{eqnarray}
as shown in Fig. \ref{Fig: QCD3}.  The symmetry 
structure translates into a constraint on the coupling of the $\rho$ to 
the pions.  

With $\kappa =0$, this theory has become a two site, two link, 
``theory space'' model.  The $\rho$ cuts off the quadratic divergence to the 
charged pion mass, just as in traditional Little Higgs theories, where 
vector bosons come in to cut off the gauge quadratic divergences to the 
Higgs boson mass.
%
The possibility that $\rho$'s could cut-off quadratic divergences to 
pseudo-Goldstone bosons was previously noted, see \cite{PionMass}.  

\subsection{Higher Modes}
\label{Sec: Higher Modes}

So far we have limited discussion to the $\rho$ meson, the lightest 
vector resonance.  It is natural 
to attempt to extend the methods employed above to more massive spin-one
resonances, such as the $a_1$.   However, our technology crucially relies 
on the lightness of the modes relative to the scale of strong
coupling.  As the resonances become heavier, 
the constraint of gauge invariance on their interactions is weakened, 
and the validity of the effective Lagrangian becomes more precarious.

In QCD the $a_1$ might already be too massive to be 
well-described by hidden local symmetry, other theories with strong
dynamics may have a wider range of states to which hidden local symmetry
can be applied.  For instance, the number of light 
resonances (those with mass $\lsim 4 \pi f_{\pi}$) scales with the number
of colors of the confining theory, $N_c$; 
so for large $N_c$ QCD, there are many light vector resonances 
and therefore more modes can be faithfully studied.
Likewise, in applications to EWSB, the $N_c$ of the UV completion may be 
larger than three, allowing for more light copies of the $\rho$ and 
$a_1$ to be described within an effective theory.

In \cite{Son:2003et,Chivukula}, an infinite number of sites was 
considered to model QCD.  However, as the number of sites increases, to keep 
the interactions of the $\rho$'s from becoming weak, the 
gauge coupling at each individual site must increase.
The result, as shown in \cite{Locality}, is that if too many sites 
are used in the effective theory, the radiative corrections to operators
non-local in the deconstructed dimensions grow exponentially large except
in special supersymmetric examples.  
Since locality is crucial to interpreting the theory as extra-dimensional,
the large non-local radiative corrections to the effective action calls into 
question the whole extra-dimensional interpretation.  In QCD, the physical
coupling of the $\rho$ is strong.  At best a few sites can be included
while avoiding this pitfall.  If the 
physical coupling of the $\rho$ were weaker, more sites could be consistently
included without encountering this difficulty.

Despite the caveats enumerated here, we will first study the theory space
representation of the $a_{1}$ in QCD.  Our ultimate goal is studying 
light resonances in a more 
general setting, but the $a_1$ in QCD gives a familiar example.  

\subsubsection{The $a_{1}$ in QCD and the Generalized Vector Limit}

\begin{figure}
\begin{center}
\epsfig{file=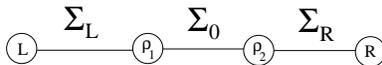, width=2.0in}
\caption{
\label{Fig: QCDA1}
Chiral Lagrangian with the lightest $a_1$ modeled.
}
\end{center}
\end{figure}

In QCD the heavier spin-one resonances  can be modeled by 
incorporating new gauge symmetries in the middle of the 
theory space.  This construction is equivalent to the construction in
\cite{HiddenLocal} and is much as \cite{Son:2003et} attempted in a more ambitious manner.

With an additional strongly gauged field, we can model the interactions of 
the $a_{1}$ (See Fig.~\ref{Fig: QCDA1}).  The Lagrangian becomes
\begin{eqnarray}
\LL &=&  -\half \Tr \rho_{1\,\mu\nu}^2 -\half \Tr \rho_{2\,\mu\nu}^2+
\frac{f^2}{4} 
\Big( 
|\Tr D_\mu \Sigma_L|^2 + c^2 |\Tr D_\mu \Sigma_0|^2 
+ |\Tr D_\mu \Sigma_R|^2 \nonumber \\
&& +\kappa |\Tr D_{\mu} \Sigma_L \Sigma_0|^2
+ \kappa |\Tr D_{\mu} \Sigma_0 \Sigma_R|^2
+\kappa' |\Tr \partial_{\mu} \Sigma_L\Sigma_0 \Sigma_R|^2
\Big),
\end{eqnarray}
where we are neglecting $U(1)_{EM}$ and also the various 
kinetic mixings for simplicity. 
Both $\rho_1$ and $\rho_2$ have the same gauge coupling, $\tilde{g}_\rho$.
There is another parameter, $c$, that is {\it a priori} undetermined.
The linearized fluctuations of the fields are given by
\begin{eqnarray}
\Sigma_L = \exp\Big(\frac{2i}{f} \pi_L\Big),
\hspace{0.5in}
\Sigma_0 = \exp\Big(\frac{2i}{cf} \pi_0\Big),
\hspace{0.5in}
\Sigma_R = \exp\Big(\frac{2i}{f} \pi_R\Big).
\end{eqnarray}
When modeling the $\rho$, we had to introduce a single non-locality 
parameter, $\kappa$.  Here we have a pair: $\kappa$ and $\kappa^{\prime}$.  
A parity transformation interchanges $SU(2)_L \leftrightarrow SU(2)_R$.  
Under this parity, there is an even field, the $\rho$, and and odd field, 
the $a_1$.  Note that the two terms with 
coefficient $\kappa$ are set equal by this parity.  

The masses of the vector mesons are
\begin{eqnarray}
\label{Eqn:VecMasses}
m^2_\rho = \frac{1 +\kappa}{4} g^2_\rho f^2,
\hspace{0.5in}
m^2_{a_1} = \frac{1 +2 c^2 +\kappa}{4} g^2_\rho f^2.
\end{eqnarray}
Note that the $a_1$ is always heavier than the $\rho$.
Setting $c=1$ and $\kappa=\kappa'=0$, 
the spectrum in the vector limit is
$m_\rho^2 = g_\rho^2 f_\pi^2$ and $m_{a_1} = \sqrt{3} m_\rho$.
While this prediction does differ from the value predicted by QCD sum rules,
$m_{a_1} = \sqrt{2} m_\rho$, it is not that far from the the 
experimental relation $m_{a_1}/m_{\rho} \simeq 1.65$. 

In the limit where all of the gauge couplings, $\kappa$ and $\kappa'$ 
vanish, there are enhanced symmetries much like those
in Georgi's vector limit.  When the generalized vector limit holds, 
the global symmetry of the theory is
\begin{eqnarray}
SU(2)_L\times SU(2)_{\rho_1} \times SU(2)_{\rho_2}\times SU(2)_R \rightarrow SU(2)_L^3\times SU(2)_R^3.
\end{eqnarray} 
Restoring the gauge couplings while keeping $\kappa$ set to zero
gives a {\it finite} Coleman-Weinberg potential because it requires
$e_0$, $g_{\rho_1}$, {\it and} $g_{\rho_2}$ couplings to
communicate sufficient chiral symmetry breaking to the effective
Lagrangian:
\begin{eqnarray}
\delta m^2_{\pi^\pm} \sim
\frac{3 e^2}{16 \pi^2} m_\rho^2\log \frac{m_{a_1}^2}{m_\rho^2}.
\end{eqnarray}
In this limit, the $\rho$ and the $a_{1}$ cut off both the
quadratic and logarithmic divergences from gauge interactions.

\subsection{Effects of Higher Modes on the Vector Limit}
\label{Sec: HigherModesKappa}

We have argued that the $\kappa \rightarrow 0$ limit is of particular interest,
partly because the quadratic divergences vanish in this limit.  We would like 
to understand how likely this limit is to obtain.  
To do this, we must understand how higher modes can affect the low-energy 
Lagrangian.  We will see that even if a theory is fundamentally local,
integrating out the heavy modes can make the effective theory for the 
lightest mode appear non-local.  

\begin{figure}
\begin{center}
\epsfig{file=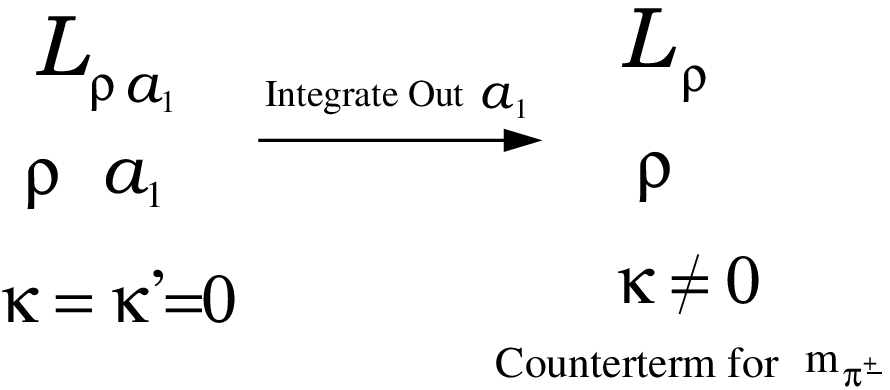, width=2.5in}
\caption{
\label{Fig: InducedKappa}
The effect of the $a_1$ on the $\pi-\rho$ Lagrangian.
The $\pi-\rho-a_1$ theory is in the generalized vector
limit where all non-local couplings are set to zero.
The radiatively induced mass for $\pi^\pm$ is finite
at one loop.
Integrating out the $a_1$ at tree level induces a $\kappa$ term,
at loop level a compensating
counter-term is induced for the $\pi^\pm$ mass.
}
\end{center}
\end{figure}

This is not too surprising.  For example, a similar phenomenon occurs in a 5D  gauge theory on a circle.  
We can Kaluza-Klein (KK) decompose the vector boson into its ladder 
spectrum, and form an effective theory by classically integrating out
all but the lightest vector boson.
In this truncated theory, a computation of the mass for the Wilson loop
operator yields an answer quadratically sensitive to the cut-off -- an answer
far larger than the finite one found by a calculation in the 
original 5D theory.  Locality in the fifth dimension prohibits 
ultraviolet contributions to the Wilson line operator.  Truncating the 
theory corresponds to placing a hard momentum cut-off in the fifth 
dimension; when Fourier transformed back to position space, this 
cut-off induces $x_5^{-1}$ correlation functions.  
Truncating the theory induces non-locality in the fifth dimension; 
the result is the Wilson loop operator no longer has exponentially small
sensitivity to the cutoff.  Properly integrating out the
tower of KK modes (at the quantum level) induces a counter-term for
the Wilson line operator that cancels against the quadratic
divergence induced by the interactions of the lightest mode.

To demonstrate how this phenomena occurs in QCD we can return to the 
formulae of the previous section and classically integrate out 
the $a_1$ resonance.  Even when starting in the
in the vector limit, a significant deviation
from $\kappa=0$ is induced.  The Lagrangian of a local 
$\rho -a_1- \pi$ theory is given by
\begin{eqnarray}
L_{\rho\,a_1}
= \frac{f^2}{4} \Big(
|D_{\mu} \Sigma_L|^2 + c^2 |D_{\mu} \Sigma_0|^2 + |D_{\mu} \Sigma_R|^2\Big),
\end{eqnarray}
with gauge couplings $\tilde{g}_\rho$ for both of the vector bosons.  
We start by normalizing $f_\pi$ through a calculation similar to the one in
Sec. \ref{Sec: QCD Rho}, which gives \cite{Son:2003et}
\begin{eqnarray}
f_\pi =\frac{f}{\sqrt{2 + c^{-2}}}.
\end{eqnarray}
The two mass eigenvalues are
\begin{eqnarray}
m^2_\rho =   \frac{\tilde{g}_\rho^2 f^2}{4} =  (2+c^{-2})\frac{\tilde{g}_\rho^2 f_\pi^2}{4}
\hspace{0.4in}
m^2_{a_1} =  \frac{(1+2c^2) \tilde{g}_\rho^2 f^2}{4}= (2+c^{-2}) \frac{(1+ 2c^2) \tilde{g}_\rho^2 f_\pi^2}{4}.
\end{eqnarray}
To relate $\tilde{g}_\rho$ to $g_{\rho}$ we consider the coupling of the 
$\rho$ to the $\pi$ isospin current.  We find  
\begin{eqnarray}
g = \frac{1+  c^{-2}}{2+ c^{-2}} \sqrt{2} \tilde{g}_\rho .
\end{eqnarray}
In the non-vector-limit theory with just the $\rho$, the mass of the $\rho$ in terms
of the isospin current coupling in Eq. \ref{Eq: Rhomasses}  and Eq. \ref{Eq: rho Current Int} 
\begin{eqnarray}
m_\rho^2 =  (1+\kappa) g^2 f^2_\pi 
\end{eqnarray}
where as in the vector-limit theory with the $a_1$ the mass of the $\rho$ is given by
\begin{eqnarray}
m_\rho^2 = \frac{(1+ \half c^{-2})^3}{(1+ c^{-2})^2} g^2 f_\pi^2
\end{eqnarray}
keeping the physical quantities $f_\pi$ and $g$ fixed, we can vary $c$ and observe
how $\kappa$ changes
\begin{eqnarray}
\kappa=  \frac{(1+ \half c^{-2})^3}{(1+ c^{-2})^2} -1 = -\left(\frac{c^{-2}}{2}\right)\left(\frac{ 1+ \half c^{-2} - \frac{1}{4}c^{-4}}{(1+ c^{-2})^2} \right)
\end{eqnarray}
There are two places where the non-locality becomes small, at $c^{-2}=0$ or
at $c^{-2}= 1+\sqrt{5}$.   At the prior, it is obvious that the theory is local because
the middle link has been contracted away and $m_{a_1}\rightarrow \infty$, while in the later, 
it is just a cancellation.
The ratio $m_{a_1}/m_\rho$ suggests that $c^{-2}=0.85$ or $\kappa=-0.15$.

Summarizing, we started with a local theory including the 
$a_{1}$ resonance, but 
after integrating out the $a_1$, the effective theory that contains only the $\rho$ is apparently
non-local (see Fig. \ref{Fig: InducedKappa}). This deviation from $\kappa =0$ occurs because
the $\rho$ and $\pi$ mix with the $a_{1}$.  This deforms the low energy effective 
action of the $\rho$--$\pi$ system away from a local/nearest neighbor interacting one.  
After truncating the theory, the counter-term to the $\pi$ mass can be computed in this
reduced theory.   The counter-term in this case turns out not to violate isospin because
the $a_1$ does not mix with the photon.  The induced counter-term violates 
does violate $SU(2)$ axial because the longitudinal components of the $a_1$ mix
with $\pi$.  If a $\rho'$ was included, then an isospin violating $\kappa$ parameter would be induced.

\subsection{Vector Limit Moral}
\label{Sec: VL}

In this section, we explore the meaning of 
Georgi's vector limit for QCD, and speculate on the implications for other 
theories with strong dynamics.  It is always possible to use an effective Lagrangian 
for those modes that are much lighter than the scale  of strong coupling.  For QCD, this 
means a Lagrangian for the $\pi$'s and the $\rho$ meson.  Keeping the longitudinal 
components of the $\rho$ explicitly in the
effective Lagrangian is useful for constraining their interactions when
the mass of the $\rho$ is light compared with the scale of strong coupling.
In this gauge, there is a term that breaks more chiral symmetries than the
others because it contains more fields.   In the vector limit this term 
vanishes.   In Naive Dimensional Analysis (NDA) \cite{NDA}, adding 
non-linear sigma model 
fields to any operator does not result in a suppression.  
Nevertheless, these $\Sigma$-laden
terms seem to be small experimentally.   So, NDA does not give any 
reason for the $\kappa$ parameter to 
be small; insight into the vector limit is beyond the scope of NDA.  The 
suppressed operators with  
additional $\Sigma$ fields are non-local in theory space.  One possible reason for this 
suppression is that such terms break more chiral symmetries than operators 
with fewer $\Sigma$ fields;  if there is a ``cost'' associated with the 
breaking a chiral symmetries, then it would be natural to expect a suppression
of these terms.

As shown in the previous section, if there are additional spin-one resonances
that mix with the $\rho$ such as the $a_{1}$ 
then the vector limit is typically spoiled.  
So if Georgi's vector limit is to hold, mixing with all heavier modes
should be small.  A theory space model with many sites incorporates mixing 
with many spin-one modes ($a_{1}$, $\rho^{\prime}$, $\rho^{\prime \prime}$),
even if the spectrum is truncated.    
To reproduce a model close to the vector limit, we must minimize this mixing.  
To do this, we write a model with sites and links for only those modes 
well-constrained by gauge invariance.  For QCD, this is probably only the $\rho$.  For the vector limit to hold, we must posit that these particles do
not have large mixing with the particles whose interactions
are unconstrained by gauge invariance.

In the next section we will generalize these statements to other 
strong coupling theories, 
such as UV completions of Little Higgs theories.  Before doing so, it is 
useful to recall how several quantities scale with $N_{c}$ \cite{LargeN}. 
We will use slightly non-standard scalings: we keep 
$f_\pi$ fixed as we vary $N_c$.   This is useful because $f_\pi$ will be an 
easily measurable quantity and in Little Higgs theories the
ratio of $f$ and $v$ will be kept fixed.    

The coupling of the $\rho$ scales as 
\begin{equation}
g_\rho \sim 4\pi/N_c^\half.
\end{equation}
With our results from before, we see that
\begin{eqnarray}
m_\rho \propto g_\rho f_\pi \sim \frac{ 4\pi f_\pi}{N_c^\half}.
\end{eqnarray}
If use $m_\rho$ as the cut-off of the low energy theory,
then we recover the standard large $N_c$ relation that
\begin{eqnarray}
\frac{\Lambda}{f_\pi} \simeq \frac{ 4\pi}{N_c^\half}.
\end{eqnarray}
In general, we will be interested in cases where the $g_{\rho}$, remains 
large, i.e.  the number of colors is not too big.  

With these large $N_{c}$ scalings in hand, we can now speculate on how
Georgi's vector limit might extend to other theories.  As before, we
should write down an effective gauge-invariant Lagrangian for the 
PNGBs and the light
vector resonances.  Because the mass of the $\rho$ scales as 
$1/\sqrt{N_{c}}$, we expect that this Lagrangian will probably 
contain more than just a single $\rho$ for non-QCD cases.  According to
NDA, there would be many non-local operators with unsuppressed 
coefficients.  On the other hand, if Georgi's vector
limit is somehow fundamental, and indeed there is a cost for breaking
chiral symmetries, we expect that these non-local operators will be 
suppressed.  This dictates that mixing with the higher 
modes (those not well described in the effective theory) is small.  

The fundamental importance of the vector limit can in principle be 
tested experimentally.  If, for example,
a EWSB involves a theory at strong coupling, one could check whether or not 
it was close to the vector limit by exploring the relationship between the
$\rho-\pi$ couplings, $m_{\rho}$, and $f_{\pi}$.

\section{The Techni-Vector Limit, Vacuum Alignment and Little Higgs Models}
\setcounter{equation}{0}
\renewcommand{\theequation}{\thesection.\arabic{equation}}
\label{Sec:TR}
Suppose that techni-$\rho$s (henceforth referred to as $\rho$s) are the 
lightest hadronic states of a multi-TeV confining gauge theory associated with EWSB.  
How do they couple to the Goldstone bosons? 
In this section, we show how to include techni-$\rho$'s in the 
low energy effective theory for composite Higgs models.  We also discuss how 
to define the analog of the vector limit.  As a first showcase for this 
formalism, we investigate the problem of vacuum alignment in these theories.  
Assuming that the vector limit does obtain, we test  some of
the results previously obtained using QCD sum rules and essentially reproduce
the major results.

We then study the implications of the vector limit for specific Little Higgs models.
Assuming there is underlying strong dynamics (at $\sim$10 TeV) which  
favors the realization of the near-vector limit, it is possible to
construct an effective theory in which some of the divergences are softened 
by these vector bosons.  This phenomenon is analogous to the way in which  
the photon contributions to the charged-pion masses were cut-off by 
interactions with the $\rho$.\footnote{One might hope that  
the contribution to the Higgs boson mass from Standard Model 
gauge interactions could be cut-off by the techni-$\rho$ \cite{Harada}.  
This softening of divergences would take place without extending the 
fundamental gauge symmetry; only the existence 
of light composite vector bosons near the vector limit is required.}  
In Secs.~\ref{Sec:Littlest} and~\ref{Sec:SU6}, we 
give two examples of Little Higgs models where inclusion of 
techni-$\rho$'s can render previously UV sensitive quantities calculable. 
First, in the $SU(6)/Sp(6)$ Little Higgs model\cite{sp6}, there
is a quadratic divergence in the Coleman-Weinberg potential.  The naive
sign suggests that the potential is unstable. However, the signs of 
quadratically divergent quantities are not necessarily to be 
trusted -- with the methods in this paper it is possible to state 
what features of the ultraviolet completion are necessary to stabilize 
the effective potential.  This is the focus of Sec.~\ref{Sec:SU6}.   
Second, in the Littlest Higgs model\cite{littlest},  
the vacuum expectation value (vev) of a electroweak triplet was 
sensitive to ultraviolet physics. We devote Sec.~\ref{Sec:Littlest} 
to addressing when the vev of the triplet may be smaller than 
naively expected, as desired by electroweak precision measurements.  
We are also able to discuss the size of the quartic coupling in this 
model.

\subsection{Coupling Techni-$\rho$'s and Vacuum Alignment}

Before proceeding to our specific examples, we discuss how to include 
techni-$\rho$'s in coset theories of electroweak symmetry breaking. 
We are particularly interested in models that have the global symmetry
structure $SU(N)/SO(N)$ or $SU(N)/Sp(N)$.  These symmetry breaking
patterns deviate from the QCD-like structure of 
$[SU(N)\times SU(N)]/SU(N)$ discussed in 
Section~\ref{Sec:QCDrho}, so it is not obvious how to
incorporate light vector resonances into the theory.  
The primary guide is that the vectors should be in a representation
of the unbroken global symmetry of the strong dynamics -- i.e. 
$SO(N)$ or $Sp(N)$.  We find it convenient to introduce the techni-$\rho$'s 
in the context of a particular problem: vacuum 
alignment \cite{VacuumAlignment}.

In theories of strong dynamics, such as the ones we are considering, 
vacuum alignment represents an important issue.  Consider a strongly 
gauged group $\GG$, with some  weakly gauged subgroups $\GG_w$.  As 
we have discussed, when $\GG$ reaches strong coupling, it generically 
breaks some global symmetries, and may also break 
some of $\GG_w$.  The issue of vacuum alignment may be succinctly 
stated as: What subgroup $\HH_w$ of $\GG_w$ is left unbroken by the 
strong dynamics?

Solving the vacuum alignment problem requires the minimization of an
effective potential.  Unfortunately, this minimization cannot be performed
without a knowledge of the bound states of the strongly coupled theory.
Reference~\cite{PeskinPreskill} utilized sum 
rules~\cite{SumRules} to provide a window on this 
non-perturbative physics.  To make 
progress, they assumed that the signs of spectral integrals could be 
determined by the lowest lying resonances.  In QCD, it is known that this 
approximation holds: the $\rho$ and the $a_{1}$ largely saturate 
the vector and axial currents.

Incorporating the techni-$\rho$'s into our effective theory, we too can 
address vacuum alignment.  We simply choose different candidates 
for $\HH_w$, and calculate the Coleman-Weinberg effective potential arising 
from gauge boson and techni-$\rho$ exchange in each case.  An instability 
in the potential means that we have chosen the wrong $\HH_w$.  

At energies beneath $m_\rho$, the PNGBs are parameterized as 
\begin{eqnarray}
\label{Eq: Goldstones I}
\Sigma = \exp(i \pi/f)\, \Sigma_0\, \exp(i \pi^T/f),
\end{eqnarray}
where $\Sigma_0$ is the symmetric (anti-symmetric) fermion condensate that
breaks $SU(N)$ to $SO(N)$ ($Sp(N)$). 
The leading action for this theory of Goldstone bosons is
\begin{eqnarray}
\LL = \frac{f^2}{4} \Tr |D_\mu \Sigma|^2,
\end{eqnarray}
where the covariant derivative $D_\mu$ is given by
\begin{eqnarray}
D_\mu \Sigma = \partial_\mu \Sigma + i  g_I W^I_\mu( T_I \Sigma + \Sigma T_I^T).
\end{eqnarray}
Here, the $W^I$ are the weakly gauged vector bosons of $\GG_{w}$ embedded 
inside the global $SU(N)$ symmetry.  We can study the question of vacuum 
alignment by examining the effective Coleman-Weinberg potential for the 
Goldstone bosons.  The gauge sector gives rise to quadratic 
divergences in the potential for the Goldstone bosons:
\begin{eqnarray}
\label{Eq:VeffGold}
V_\eff (\Sigma) \ni  \frac{3\Lambda^2 }{16 \pi^2} \sum_{I=J} \OO_{IJ}
\end{eqnarray}
with 
\begin{eqnarray}
\OO_{IJ}(\Sigma) = \frac{f^2}{4} \Tr\Big[( g_I T_I \Sigma + g_I \Sigma T_I^T) (g_J \Sigma^\dagger T_J + g_J T^T \Sigma^\dagger)\Big] .
\end{eqnarray}
At this point, we cannot say anything definitive about the stability 
of the effective potential.  The quadratic divergence indicates sensitivity 
to the ultraviolet, and its sign, central to the question of stability, is not
to be trusted.

Now we add the techni-$\rho$'s. By taking these vector mesons 
to be in the adjoint representation of the unbroken chiral symmetry,  
we can incorporate them into the theory by defining a new multiplet of 
Goldstone bosons:
\begin{eqnarray}
S \equiv \exp( i \Pi/f) \mbox{ with } \Sigma = S \Sigma_0 S^T.
\end{eqnarray}
Now $\Pi$ is an  adjoint of $SU(N)^2/SU(N)$ Goldstone bosons, 
instead of the $SU(N)/SO(N)$ or $SU(N)/Sp(N)$ Goldstone multiplet we had 
previously.  The longitudinal components of the techni-$\rho$'s have 
filled out the remainder of the multiplet.  $\Sigma$ contains the 
same set of Goldstone bosons from Eq.~\ref{Eq: Goldstones I} because the 
additional Goldstone bosons commute through $\Sigma_0$ and cancel.  Note that 
the $SU(N)^{2}$ symmetry of the Lagrangian is broken by the vacuum 
expectation value $\Sigma_{0}$.

The Lagrangian incorporating the $\rho$'s is given by
\begin{eqnarray}
\label{eqn:newL}
\LL_{T\rho} = f^2 (
\Tr |D_\mu S|^2 + \frac{\kappa}{4} \Tr|D_\mu \Sigma|^2 ).
\end{eqnarray}  
The covariant derivatives are given by
\begin{eqnarray}
\nonumber
\label{eqn:cosetDs}
D_\mu S &=& \partial_\mu S + i g^0_I W^I_\mu T_I S + i g_\rho S T_A \rho_\mu^A, \\
D_\mu \Sigma &=& \partial_\mu \Sigma + i  g^0_I W^I_\mu( T_I \Sigma + \Sigma T_I^T),
\end{eqnarray}
where $\rho^A$ are the techni-$\rho$'s and $A$ runs over the adjoint of 
$SO(N)$ or $Sp(N)$, depending on the unbroken global symmetry group.

We can now compute the Coleman-Weinberg potential for the theory
with the techni-$\rho$ mesons.  The techni-$\rho$'s mix with
the vector bosons of the weakly gauged symmetries and produce a 
logarithmic divergence.  
If $\kappa =0$, then there would be no  $\pi^2 W^2$ coupling
or $\pi^2 \rho^2$ coupling, only a $\pi^2 W \rho$ coupling.  This interaction 
cannot produce a one-loop quadratic divergence.  
The resulting Coleman-Weinberg potential is 
\begin{eqnarray}
\label{Eq:EffPotwithRho}
V_\eff(\Sigma) = \frac{3}{16\pi^2} \left(
\frac{\kappa \Lambda^2}{ \cos^{2} \theta}   
+ 2 m_\rho^2 \log{\Lambda^2} \right) \sum_{I=J}\OO_{IJ}(\Sigma) + \cdots \, ,  
\end{eqnarray} 
where we have written the result in terms of the low-energy gauge coupling, 
$g_{I} \equiv g_{I}^{0} \cos \theta$ and $m_\rho$. The ``$\cdots$'' includes a 
logarithmically divergent term, but is proportional to $\kappa$.  
The $\OO_{I}$ are the same operators as in 
Eq.~\ref{Eq:VeffGold},
there is no explicit  dependence on $S$ alone.  This should not be 
surprising -- the additional 
modes present in $S$ not contained in $\Sigma$ are eaten (exact) Goldstone 
bosons and, as such, cannot have a potential.  
To put the potential in this form, we 
have performed the sum over the $SO(N)$ or $Sp(N)$ adjoint,
and the fields arrange themselves in the form $\Sigma = S \Sigma_0 S^T$. 

Before discussing the implications of Eq.~\ref{Eq:EffPotwithRho} for vacuum 
alignment, we first discuss the symmetry structure of the techni-vector limit.
As in the two-flavor QCD example, the quadratic divergence vanishes 
due to an enhanced symmetry of the
vector limit when $\kappa=0$.  In this limit, when the weak gauge couplings,
$g_I$, and the $\rho$ gauge couplings, $g_\rho$, vanish, 
it is possible to perform {\it independent} $SU(N)\times SU(N)_\rho$ 
transformations on $S$:
\begin{eqnarray}
S \rightarrow G S G_\rho.
\end{eqnarray}
The term proportional to $\kappa$ forces $G_\rho$ to be a
$SO(N)$ or $Sp(N)$ global transformation.  


In the $\kappa=0$ limit, the sole vestige of the breaking of the global 
$SU(N)$ symmetry by the strong dynamics is the representation of the
techni-$\rho$s. They make up a multiplet of $SO(N)$ or $Sp(N)$ 
(rather than the full $SU(N)$). 
Having the lightest vector resonance communicate symmetry breaking
offers an interesting view on how global symmetries 
are dynamically broken by fermion condensates.

Now we can return to the implications of Eq.~\ref{Eq:EffPotwithRho} for vacuum
alignment.  Since the quadratic divergence vanishes, the potential is 
dominated by the logarithmically divergent piece.  To determine the 
stability of a given the vacuum alignment, one would in principle expand out 
the operators, $\OO (\Sigma)$ and see whether all PNGB's receive a 
positive (mass)$^2$.  However, we can take a shortcut to the result by 
simply noting that the operators of Eq.~\ref{Eq:EffPotwithRho}
operators are identical to those produced in the 
traditional \cite{PeskinPreskill} vacuum alignment studies.  

Our agreement with the QCD sum rules results from the $\rho$'s 
regulating the quadratic divergences.  To reverse the quadratic divergence, 
an $a_1$-like resonance should be dominant over
the $\rho$.  At this stage, we have not even incorporated such a 
mode.   In Sec. \ref{Sec: TechniHigher Modes} 
we discuss the incorporation of higher modes in strong coupling theories, 
and whether being very far from the vector limit can change this result.  

\subsection{Little Higgs Examples}
Little Higgs models often contain ultraviolet-sensitive 
quantities whose values are crucial for determining the viability 
of the theory.   Because the ultraviolet completions of these 
models are unspecified, it seems impossible to comment definitively 
on the viability of these models.  If we suppose the vector 
limit obtains, these quantities can be computed within an effective theory. 
This allows us to discuss the viability of the models in the vector limit.
We will give examples of how to apply our technology to specific 
Little Higgs models.

\subsubsection{The Littlest Higgs: $SU(5)/SO(5)$}
\label{Sec:Littlest}

The Littlest Higgs is the best known Little Higgs model and
has been studied in some depth\cite{PrecisionEWLittlest,ColliderTests}.  
It is closely related to the Georgi-Kaplan composite Higgs 
model, which has the same coset structure\cite{PGBHiggs}.  This coset 
was originally chosen to preserve custodial $SU(2)$, thereby protecting 
the $T$ parameter.   The primary difference between the Littlest Higgs and the 
Georgi-Kaplan model is in the weakly gauged part of $SU(5)$.
The Littlest Higgs model we consider gauges 
$SU(2)_1\times SU(2)_2\times U(1)_Y$ in $SU(5)$; because only a single $U(1)$ is gauged, there is an additional axion--like field beyond the Goldstone bosons 
discussed in \cite{littlest}.  The gauge symmetry is 
broken to $SU(2)_L\times U(1)_Y$ by the
condensation.  In the Georgi-Kaplan composite Higgs model, 
only the standard model $SU(2)_L\times U(1)_Y$ is gauged.  Gauging the 
additional $SU(2)$ breaks custodial $SU(2)$.  This results in a 
$SU(2)_L$ triplet, hypercharge one scalar 
acquiring a phenomenologically problematic vev -- however, it is 
also the reason that the Higgs doublet acquires a large quartic potential, 
leading to viable EWSB.  

To see how the triplet acquires a vev, consider the gauge quadratic 
divergence in the Coleman-Weinberg potential:
\begin{eqnarray}
V_\eff = \frac{3 \Lambda^2}{16 \pi^2}\Big(
 \OO_{1} +  \OO_2 +  \OO_Y \Big),
\end{eqnarray}
with 
\begin{eqnarray}
\nonumber
&&\OO_1 = \frac{g_1^2}{4}\left(\Tr \left| \phi - \frac{i\,hh^T}{2 \sqrt{2} f}\right|^2 +\cdots\right),
\hspace{0.45in}
\OO_2 = \frac{g_2^2}{4} \left(\Tr \left|\phi + \frac{i\,hh^T}{2 \sqrt{2} f}\right|^2 +\cdots\right),
\\
&&\hspace{1.4in} \OO_Y =  g'{}^2\left(|\phi|^2+ \frac{1}{4}|h|^2 - \frac{1}{24f^2} |h|^4\cdots\right),
\end{eqnarray}
where we have neglected the interactions of the axion-like $\eta$.

The triplet vev arises from the term $h^T\phi^\dagger h$ in the expansion
of these operators.  

Integrating out the massive $\phi$ field induces both a Higgs quartic coupling
and a dimension six operator that contributes to a $T$ parameter.
\begin{eqnarray}
\label{eqn:dim6T}
\LL_\eff = \lambda |h|^4 +\frac{c_T}{f^2} |h^\dagger D_\mu h|^2,
\end{eqnarray}
with
\begin{eqnarray}
\lambda= \frac{1}{16\pi^2 f^2} \frac{ g_1^2 \Lambda_1^2 g_2^2\Lambda_2^2}{g_1^2 \Lambda_1^2 + g_2^2\Lambda_2^2} ,
\hspace{0.3in}
c_T= \frac{(g_1^2 \Lambda_1^2/f - g_2^2\Lambda_2^2/f)^2}{ 8(g_1^2\Lambda_1^2 + g_2^2\Lambda_2^2 + 4g'{}^2\Lambda_Y^2)} ,
\end{eqnarray}
where we have allowed for the possibility of different cut-offs for the operators
$\OO_1$ and $\OO_2$.
Because the coefficients of the operators are quadratically 
divergent, it is not clear how to interpret this result -- for example,
one could imagine physics in the ultraviolet cutting off the divergences
differently.  For instance, one might have $\Lambda_{1,2} \propto g_{1,2}^{-1}$ 
that would cancel the induced $T$ parameter in the above formula.   

We now suppose that the vector limit obtains, and repeat the calculation.  
First, we have to incorporate the $\rho$ mesons.
The $\rho$'s form an adjoint of $SO(5)$ and decompose under 
$SU(2)_L\times SU(2)_R$ and $SU(2)_L\times U(1)_Y$ as
\begin{eqnarray}
\mathbf{10}\rightarrow 
(\mathbf{3},\mathbf{1})\oplus (\mathbf{2},\mathbf{2})\oplus (\mathbf{1},\mathbf{3})\rightarrow
\mathbf{3_0}\oplus\mathbf{1_0}\oplus \mathbf{1_1}
\oplus\mathbf{2_\half}.
\end{eqnarray}
The $\mathbf{1_0}$ and $\mathbf{1_1}$ form an $SU(2)_R$ triplet.
Although the weakly gauged sector does not respect
custodial $SU(2)$, the strong resonances do, and the strong dynamics
is identical to that of the Georgi-Kaplan composite Higgs model. 
The additional spin-one resonances do not
mediate large effects to the $T$ parameter precisely because of the custodial 
$SU(2)$ An analysis similar to the one used for the gauge sector of a Little 
Higgs model in Ref.~\cite{ChangWacker} explicitly shows this is the case.

\begin{figure}
\begin{center}
\epsfig{file=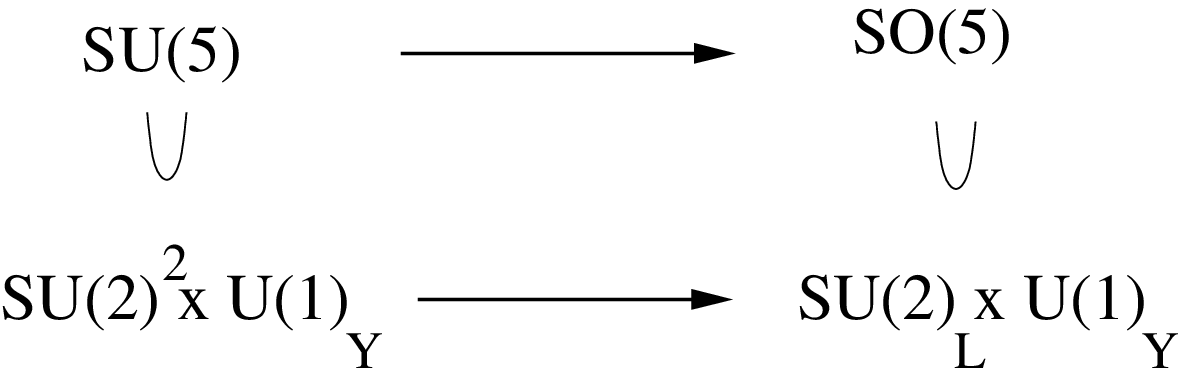, width=2.25in}
\caption{
\label{Fig: LH1}
The global and gauge symmetry structure of the $SU(5)/SO(5)$ 
composite Higgs model.   The electroweak $SU(2)_L\times U(1)_Y$
gauge sector is embedded within the global $SO(5)$. 
}

\end{center}
\end{figure}
\begin{figure}
\begin{center}
\epsfig{file=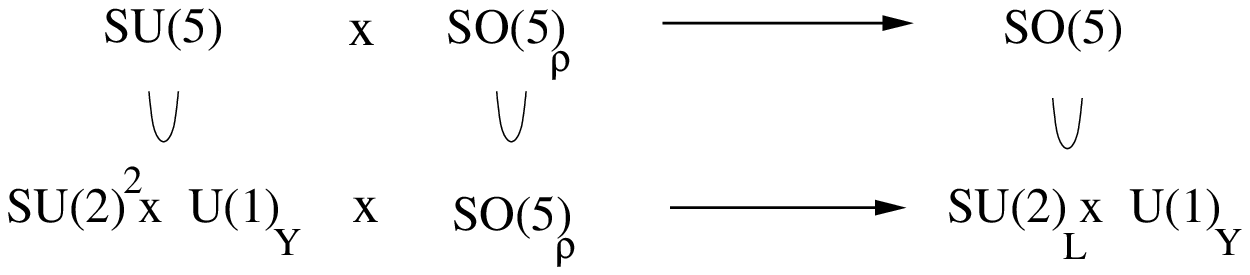, width=3.1in}\hspace{0.45in}\epsfig{file=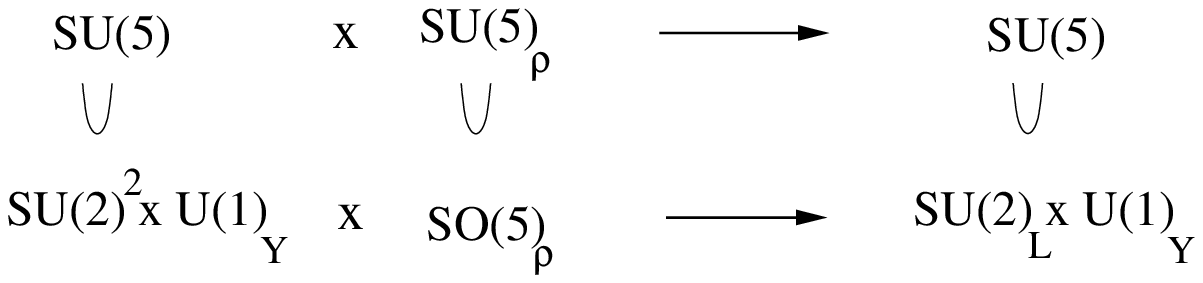, width=3.1in}
\caption{
\label{Fig: LH2}
The global and gauge symmetry structure of the $SU(5)/SO(5)$ model with a 
light techni-$\rho$.   The left diagram shows the symmetry structure of such a theory,
while the right shows the light techni-$\rho$ in the vector limit. 
}
\end{center}
\end{figure}


As in the previous section, in the vector limit the 
techni-$\rho$s cut off the gauge quadratic  divergence: 
\begin{eqnarray}
\label{eqn:VLLittlest}
V_\eff = \frac{3 m_\rho^2}{8 \pi^2} \log \Lambda^2 \Big(
\OO_{1} + \OO_2 +  \OO_Y \Big).
\end{eqnarray}
Note that this does not alter the naive prediction for the 
triplet vev because it provides the same cut-off for the quadratic 
divergences to $\OO_1$ and $\OO_2$.    The induced dimension six operator
is identical to that of Eq.~\ref{eqn:dim6T}, but with 
$\Lambda^2 \rightarrow m^2_{\rho}\log \Lambda^2$.   

The two $SU(2)$ gauge couplings are related to the Standard Model $SU(2)$ coupling
by $g^{-2}= g_1^{-2}+g_2^{-2}$.  We define the mixing angle $\tan \theta \equiv g_1/g_2$
as the new low energy parameter.
In the vector limit, the triplet mass is
\begin{eqnarray}
m^2_\phi =\frac{ m_\rho^2}{8\pi^2}  \frac{3 g^2}{\sin^2 2 \theta} 
(1+ \tan^2\theta_{\text{w}} \sin^22\theta)
= 6 m^2_{W'} (1 + \tan^2\theta_{\text{w}} \sin^22 \theta)\frac{ m_\rho^2}{\Lambda^2}.
\end{eqnarray}
where $m_{W'}$ is the mass of the TeV scale $W'$ vector boson.
Then the triplet vev is given by 
\begin{eqnarray}
\label{Eq:TripVev}
\langle \phi\rangle =  \frac{ v^2 \cos 2 \theta}{4\sqrt{2}f (1 + \tan^2\theta_{\text{w}} \sin^2 2 \theta)}.
\end{eqnarray}
This vacuum expectation value is independent of the
$N_c$ in the ultraviolet completion.  It only depends on the 
ratio of the two $SU(2)$ gauge couplings; the vev vanishes when the 
two couplings are equal.   For reference, barring cancellations, the 
experimental limit on a triplet vev is 3 GeV.  Thus, Eq.~\ref{Eq:TripVev} 
implies strong constraints for $f \lsim 2.5 \TeV$ unless there are
cancellations.

The Higgs quartic coupling results from 
integrating out the triplet field.  In the vector limit, it is given by
\begin{eqnarray}
\label{Eqn: summaryquartic}
\lambda = \frac{ 3 g^2m_\rho^2 }{4 \Lambda^2}\frac{1+ \frac{2}{3}\tan^2\theta_{\text{w}} - \frac{1}{3} \tan^4\theta_{\text{w}} \sin^2 2\theta}{1 + \tan^2\theta_{\text{w}} \sin^2 2\theta} 
\hspace{0.2in}
\Rightarrow 
\hspace{0.2in}
m_h \simeq \sqrt{6} m_W(1 + \frac{1}{3} \tan^2\theta_{\text{w}})\frac{m_\rho}{\Lambda}
\end{eqnarray}
where we have dropped terms proportional to $\sin^22\theta$ in the expression for $m_h$
because $\theta$ needs to be small from precision electroweak constraints \cite{PrecisionEWLittlest, ChangWacker,PrecisionEWSp6}.
If the Littlest Higgs theory is to 
produce an adequately heavy Higgs, this points to a  UV completion
with heavy techni-$\rho$s.
This prediction for the Higgs boson mass is subject to large corrections from the top quark Yukawa
coupling.   

Summarizing, the triplet vev is independent of the any details about the
$\rho$s as they only provide a universal cut-off for the gauge quadratic divergences.
The Higgs mass and triplet mass are proportional to the mass of the lightest $\rho$.
As discussed in Sec. \ref{Sec: VL}, the $\rho$ mass scales the number of colors in the
confining theory, $m_\rho \simeq 4 \pi f/N_c^\half$.  Thus, a heavy $\rho$ points to a 
small $N_{c}$ UV completion.
As a side note, the $\rho$ can mediate an $S$ parameter
\begin{eqnarray}
\delta S \propto \frac{ v^2}{m_\rho^2} \sim N_c \frac{ v^2}{\Lambda} .
\end{eqnarray}
where $\Lambda = 4\pi f$.   This indicates heavy techni-$\rho$s, roughly
corresponding to a small $N_c$ UV completion.

\subsubsection{The Vacuum of $SU(6)/Sp(6)$}
\label{Sec:SU6}

The Little Higgs model $SU(6)/Sp(6)$  \cite{sp6} has garnered 
attention\cite{PrecisionEWSp6} because it possesses many of the 
requisite properties to be a minimal Little Higgs model\footnote{
Note that our definition of $f^2$ is a factor of 2 greater than in \cite{PrecisionEWSp6}.
}.  However, 
there is one potential problem with the model --  the Coleman-Weinberg 
potential is unbounded if one takes the naive sign.   This is worrisome, 
but not necessarily fatal; the Coleman-Weinberg potential is dominated 
by cut-off scale contributions which could potentially reverse the
 naive sign.  
In this section we assume that the lightest $\rho$ is in the vector limit.
Then the quadratic divergence vanishes, and the effective potential 
becomes calculable.  Under these assumptions, we find that the potential
remains unstable.

The $SU(2)_1\times SU(2)_2\times U(1)_Y$ quadratically divergent
contribution to the scalar potential is
\begin{eqnarray}
V_\eff =  \frac{ 3 \Lambda^2}{16\pi^2} \Big(
 \OO_{1} + \OO_2 +  \OO_Y \Big)
\end{eqnarray}
with 
\begin{eqnarray}
\nonumber
&&\OO_1 = - \frac{g_1^2}{4}\left(\left|\eta + \frac{i\,h_1^\dagger h_2}{2\sqrt{2}f}\right|^2 + \cdots\right)
\hspace{0.5in}
\OO_2 = - \frac{g_2^2}{4}\left(\left|\eta - \frac{i\,h_1^\dagger h_2}{2\sqrt{2}f}\right|^2 + \cdots\right)\\
&&
\hspace{1.5in}
\OO_Y = \frac{g'{}^2}{4}\left(\left(|h_1|^2 +|h_2|^2\right) + \cdots \right).
\end{eqnarray}
In the vector limit, the techni-$\rho$'s 
cut-off the gauge quadratic divergence and {\it do not reverse the naive sign}:
\begin{eqnarray}
V_\eff = \frac{3  m_\rho^2}{16 \pi^2}\log \Lambda^2 \left(
 \OO_{1} +  \OO_2 +  \OO_Y \right)\propto - (g_1^2+g_2^2) m_\rho^2|\eta|^2+\cdots.
\end{eqnarray}
This indicates that the vacuum, $\Sigma_0$ is unstable in the vector limit.    
In the vector limit, the gauge sector destabilizes the vacuum.  Unless there are other contributions 
to these operators,  this theory is unstable.   The vacuum preferred by the gauge sector is the one that preserves
\begin{eqnarray}
SU(2)_1\times SU(2)_2\times U(1)_Y \subset Sp(6).
\end{eqnarray}
One possibility is that the theory could be far away from the vector limit with $\kappa<0$ 
and reverses the naive sign of the quadratic divergence.  Of course the sign can not be trusted 
and the theory should be matched on to one that includes a new techni-$\rho$' that regulates this
quadratic divergence.  We briefly discuss this in  Sec. \ref{Sec: TechniHigher Modes}, where we discuss the incorporation of higher resonances.

\subsection{Higher Modes in Coset Models}
\label{Sec: TechniHigher Modes}

\begin{figure}
\begin{center}
\epsfig{file=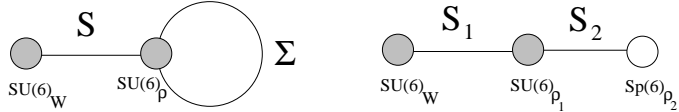, width=3.5in}
\caption{
\label{Fig: TechniA1}
To extend $SU(6)/Sp(6)$ beyond the lightest $\rho$ to include the
$a_1$ one must enlarge the gauge structure.  To include the additional
$a_1$, promote the gauged $Sp(6)_\rho$ to $SU(6)_\rho$, as depicted
in the left figure.   The following $\rho'$ can be included by adding another
$Sp(6)_\rho$, as shown in the right figure.
} 
\end{center}
\end{figure}

Given the discussion on QCD,  we expect that the inclusion of an 
additional gauge
field should allow us to model ``techni-$a_{1}$'s.''  After adding this gauge 
group, the first issue is how to distinguish between the $a_1$ and $\rho$. 
This is easily solved in the coset models we have considered because they 
possess a parity that reverses the broken generators, $X$, while leaving 
invariant the unbroken ones, $T$.   We identify the techni-$\rho$ with the 
state of even parity.
The typical theory space diagram after the inclusion of the $a_{1}$ 
is shown in Fig.~\ref{Fig: TechniA1}.  
The way to incorporate higher modes into coset theories is the natural
extension of the way one would do it QCD.  In this section
we will use $SU(6)/Sp(6)$ as an example; the extension to
other coset models is straight-forward.   

To model resonances beyond the $\rho$ in these non-QCD-like theories
it is necessary to enlarge the gauge structure.   To include the $a_1$
we promote the strongly gauged $Sp(6)_\rho$ to $SU(6)_\rho$ 
and include two non-linear sigma model fields: one a bi-fundamental 
under $SU(6)_w$ and $SU(6)_\rho$, $S$,and 
an anti-symmetric tensor under $SU(6)_\rho$, $\Sigma$.   This action 
contains a single non-local operator:
\begin{eqnarray}
\LL &=&  -\half \Tr P_\munu^2  
+  f^2\Big( \Tr |D S|^2 + c^2 \Tr |D \Sigma|^2 
+ \kappa \Tr| D S\Sigma|^2 
\Big)
\label{Eq: SU6 a1}\,.
\end{eqnarray}
As in the case with only the $\rho$ incorporated, $\kappa$ alone determines
the one-loop quadratically divergent counter-term for the effective potential.
For $\kappa \neq 0$, there is a quadratic divergence, and we cannot
reliably calculate the vacuum alignment using our techniques.  For 
$\kappa = 0$, the vector limit, the contribution to the effective potential 
from the $\rho$'s dominate over the contribution from the $a_{1}$, and 
we have the naive vacuum alignment.  Thus, we need to introduce resonances
beyond the $\rho$ and the $a_1$ if we wish to study possible modifications of 
vacuum alignment.

The next simplest theory includes the $\rho$, $a_1$ and the $\rho'$.
This theory has a strongly gauged $SU(6)_\rho$ and $Sp(6)_\rho$ and 
is depicted in Fig. \ref{Fig: TechniA1}. There are several 
non-local interactions.  The Lagrangian is given by
\begin{eqnarray}
\nonumber
\LL &=&  -\half \Tr W_\munu^{2} -\half \Tr P_\munu^2  - \half \rho_\munu^2
+  f^2\Big( \Tr |D S_1|^2 + c^2 \Tr |D S_2|^2 \\
&& \hspace{0.5in}
+ \kappa_1 \Tr| D S_1 S_2|^2 + \frac{\kappa_2}{4} \Tr| D S_2 \Sigma_0 S_2^T|^2 
+ \frac{\kappa'}{4} \Tr| D S_1 S_2 \Sigma_0 S_2^TS_1^T|^2 
\Big)\label{su6}\,,
\end{eqnarray}
where we define $S_j=e^{i\,\pi_j/f}$.

\subsubsection{Salvaging $SU(6)/Sp(6)$ Away from the Vector Limit?}
\label{Sec: Saving SU6}

In Sec. \ref{Sec:SU6} we showed that the vector limit gave the
wrong vacuum for the $SU(6)/Sp(6)$ to be a Little Higgs theory.  However, the vector
limit can be altered by mixing with heavier modes as we saw
in Sec. \ref{Sec: HigherModesKappa}.  Given this, can we construct
a theory incorporating higher modes, sufficiently far
away from the single $\rho$ vector limit of Sec. \ref{Sec:SU6},
so that the theory has the vacuum required to be a Little
Higgs model?  At the same time, we wish to maintain a good description
in terms of an effective Lagrangian of PNGB's and light resonances
without quadratic divergences in the Coleman-Weinberg potential.
In the traditional sum rule picture, it is assumed 
that the vacuum is largely determined by the lowest resonances, which
saturate spectral functions.  We would like to understand the robustness of
this result.

By modeling the $a_1$ resonances in the way we discussed here, it is possible to provide 
some more explicit understanding of how the 
vacuum could be modified from its naive alignment.  
Since the effective potential for the PNGB's of the theory
is produced by non-trivial mixing between spin-1 states 
after the inclusion of spontaneous and explicit symmetry breaking terms,
it depends not only on the actual masses of the $a_1$ and $\rho$ fields, but also 
on the mixing with the low energy states.
Most importantly, $a_1$'s and $\rho$'s contribute to such potential with
opposite sign, so that it is in principle possible to reverse the naive 
sign of the Coleman-Weinberg potential by increasing the realtive importance of the 
$a_{1}$'s.

We performed extensive studies trying to reverse vacuum alignment,
focusing in particular on cases in which only local operators are
allowed at the effective Lagrangian level, in such a way as to 
soften, or even remove, UV cut-off dependences in the Coleman-Weinberg 
potential.
We found that, indeed, thanks to the complicated form of mixing
terms in the spin-1 field mass matrices,
it appears possible to stabilize the vacuum, but only at the price of
using negative values for the $\kappa$ factors, and somewhat large ratios 
between the gauge couplings.   This implies that the 
vacuum can be stabilized only in the parameter space region where the 
theory starts to become strongly coupled: some of the states have masses 
close to the natural cut-off
of the theory, and the validity of the effective field theory description 
becomes questionable.  While not conclusive,  this study does indicate that 
vacuum alignment seems to be more robust than what one might have 
naively expected. 

\section{Unitarity and Strong Coupling}
\setcounter{equation}{0}
\renewcommand{\theequation}{\thesection.\arabic{equation}}
\label{Sec:unitarity}

Little Higgs models are non-renormalizable non-linear sigma models and
so must become strongly coupled.  At some scale, $\Lambda$,  the 
low energy description of the theory becomes inadequate, and new physics 
(or strong coupling) sets in.  Naive dimensional analysis (NDA) \cite{NDA} 
gives a cut-off of $\Lambda \sim 4 \pi f$.  There are many simple 
refinements of the NDA estimate.  One of the most common is the large 
$N_c$ refinement, which estimates $\Lambda \sim 4 \pi f/N_c^\half$.  
A similar result applies 
for a large number of fermions, $N_f$: $\Lambda \sim 4 \pi/N_f^\half$.

An alternate approach is to do a partial-wave unitarity 
analysis, as done for Little Higgs models in \cite{Chang:2003vs}.  
One examines the amplitudes of the 
Goldstone bosons scattering, and finds 
that the scattering becomes strong at 
\begin{equation}
\Lambda_U \sim \sqrt{4\pi} f/\NN^{\frac{1}{4}},
\end{equation}
where $\NN$ is the number of Goldstone bosons. $\NN$ 
roughly scales as $N_f^2$,  so this result for $\Lambda$ roughly matches 
the large $N_f$ refinement of NDA, up to a difference of $\sqrt{2\pi}$.  
In this large $N_{f}$ limit, the difference is due to 
conservatism -- the partial wave analysis is more conservative than NDA.  
Note, chiral symmetry breaking cannot produce an arbitrarily large number of 
PNGBs for a fixed $N_c$: the number of PNGBs roughly scales as 
$\NN\sim N_f^2$, and at sufficiently large $N_f$ the confining
theory becomes asymptotically non-free.

Both the NDA and the partial wave analyses attempt to give an indication of 
where new physics modifies the scattering behavior of the Goldstone bosons.  
It is natural to inquire what form this new physics takes.  
There are some arguments that vector mesons are responsible
for the unitarization of $\pi^2\rightarrow \pi^2$ scattering at intermediate
energies, i.e. $\rho$ mesons soften the $\pi$ scattering.  In this section we 
explore whether this possibility obtains.   In fact, we find that 
incorporating the $\rho$ mesons does {\it not} result 
in a parametric rise in scale where perturbative unitarity is lost.  
There are special values of $\kappa$ where the leading $\pi^4$ interactions
vanish but the $\xi^2\pi^2$ and $\xi^4$ interactions do not disappear.  

While the $\rho$ does not provide a 
panacea, it can conceivably give  a temporary postponement of the
scale of perturbative unitarity violation.  
Recall that the coupling of the $\rho$ 
scales as $g_\rho \sim 4\pi/N_c^\half$, so the mass of the $\rho$ scales 
as 
\begin{eqnarray}
m_\rho \sim \frac{4\pi f}{N_c^\half}.
\end{eqnarray}
Note, this coincides with the large $N_c$ cut-off.  In this large $N_{c}$
case, one might visualize a series of vector resonances, starting with the 
lowest $\rho$ states, serving
to postpone unitarity violation.  This is not
dissimilar to the ``Higgs''-less theories \cite{Higgsless}
or Randall-Sundrum I models 
that provide a window of unitarization mediated by Kaluza-Klein modes.   
For small $N_c$, where the vector resonances are
heavy, it seems clear that the techni-$\rho$s are not responsible for 
unitarizing the scattering.   In this case, incorporating a 
broad $\sigma$-like resonances seems more reasonable.  We discuss this 
in more detail in 
Sec. \ref{Sec: Sigma Resonance}.  In this case, we interpret the partial wave 
unitarity scale as where the broad $\sigma$ resonance appears.  

\subsection{$\rho$'s and Unitarization}

We now address the onset of strong coupling in the presence of the  
$\rho$'s.
Therefore, we expect that the scale of strong pion scattering is
modified by the addition of the $\rho$ mesons to the chiral Lagrangian.
Unfortunately, the energy region where the $\rho$ is important for
scattering is not in the equivalence region where $E\gg m_\rho$, meaning
that the transverse components of the $\rho$ matter for scattering.   
This
complicates the results.  For simplicity
we ignore this technicality and only consider the
longitudinal component--our results will be most accurate in theories
where the $\rho$'s are light.

To explore strong coupling, we first expand the original $\pi$  
Lagrangian in
Eq.~\ref{Eq: QCD1} to quartic order in the fields to find how the
interactions behave without the influence of the $\rho$:
\begin{eqnarray}
\label{Eq: QCD1 Quartic}
\LL_\eff \supset
\frac{2}{3 f_\pi^2} \Gamma^{(4)}(\pi,  D\pi)
= \frac{2}{3 f_\pi^2} \Gamma^{(4)}_{abcd} \pi^a\pi^b D\pi^c  D\pi^d,
\end{eqnarray}
where
\begin{eqnarray}
\Gamma^{(4)}_{abcd} = \Tr
\tau_a \tau_b \tau_c \tau_d
- \Tr \tau_a \tau_c \tau_b \tau_d = \frac{1}{4} \Big( \delta_{ab}  
\delta_{cd} - \delta_{ac}\delta_{bd}\Big).
\end{eqnarray}
Similarly, the chiral Lagrangian containing the $\rho$ can also be  
expanded to
quartic order.  One must take care to canonically normalize
$\xi$, the longitudinal component of the $\rho$. In this case, the
quartic interactions are
\begin{eqnarray}
\nonumber
\LL &\supset&
\frac{2 N_\pi^2}{3  f_\pi^2}\Big[
\Gamma^{(4)}\left((N_\pi^\inv \pi+N_\xi^\inv \xi),D(N_\pi^\inv \pi  
+N_\xi^\inv \xi)\right)\\
\nonumber
&&\hspace{0.4in}
+\Gamma^{(4)}\left((N_\pi^\inv \pi-N_\xi^\inv \xi),D(N_\pi^\inv \pi  
-N_\xi^\inv \xi)\right)\\
&&\hspace{0.4in}
+\frac{\kappa}{2}\Gamma^{(4)}(2N_\pi^\inv\pi,2N_\pi^\inv D\pi)\Big].
\end{eqnarray}
This effective Lagrangian will break down due to strong coupling at  energies
not drastically different from $4 \pi f_{\pi}$, even with the  incorporation
of the $\rho$'s into the Lagrangian.

In the vector limit the scattering simplifies significantly: the final  term
in the Lagrangian is absent and $N_{\pi} = N_{\xi}$. In the vector  limit,
the lowest scale of unitarity violation involves the scattering of the state
\begin{eqnarray}
|\phi\rangle= \frac{1}{\sqrt{2}} ( |\pi\rangle \pm |\xi\rangle).
\end{eqnarray}
This corresponds to a state localized in $\pi_L$ or $\pi_R$.
This is clear from the geometric picture where the states with
lowest scale of strong coupling are localized wave packets,
rather than mass eigenstate.  As localized object, they probe short  distances,
and thus the highest energies.  As we move away from the
vector limit, the most strongly coupled state
is more  $\pi$-like when $\kappa>0$ and more $\xi$-like when $\kappa<0$.

The scale of unitarity violation for a localized states is related to the $f$ of the
adjacent link.  As more modes are incorporated, the $f$ associated with
each link increases relative to $f_\pi$:
\begin{eqnarray}
f = n_S^\half f_\pi,
\end{eqnarray}
where $n_{S}$ is the number of sites.  This increases the separation
between $f_{\pi}$ and $f$, showing an improvement in the scattering behavior.
The introduction of each vector resonance allows the temporary   postponement
of the scale of unitarity violation, similar to Higgsless theories of  EWSB
\cite{Higgsless}, where the new KK modes postpone the breakdown of 
unitarity in longitudinal $W W$ scattering.

In the generalized vector limit with $n_S$ sites,  the scale of  
unitarity violation simply scales as
\begin{eqnarray}
\Lambda_U \propto n_S^\half f_\pi
\end{eqnarray}
because  the $\Gamma^{(4)}$ tensors are block diagonal and different
sets of pions don't interact at leading order.
On the other hand, if the theory is significantly away from the vector
limit, the number of PNGBs in the final state increases as $n_S$, thus
{\it lowering} the scale where unitarity violation would occur.  The  
scale
of unitarity violation scales as
\begin{eqnarray}
\Lambda_U \propto n_S^{\frac{1}{4}} f_\pi .
\end{eqnarray}
Thus the vector limit appears to help stave off unitarity violation  making the
theory healthy over a larger energy regime.

\subsection{The $\sigma$ Resonance}
\label{Sec: Sigma Resonance}

There are several broad light isospin singlet scalar resonances in QCD,
typically called $f_0$ resonances.
The first $f_0$ lies in the 500 MeV range with a width
roughly as big as its mass.  These states are capable of unitarizing
$\pi\pi$ scattering exactly as the Higgs boson does -- by being the
fluctuation of the vacuum expectation value.

In QCD $N_c$ and $N_f$ are the same, and the $\rho$ and $f_0$ are
roughly degenerate. In this section we argue that in the limit where
there are a large number of pions (corresponding to a
large $N_{f}$), the $f_0$
resonance becomes light and is responsible for
unitarizing the Nambu-Goldstone boson scattering.
Typically  several scalar resonances are necessary
to completely unitarize $\pi\pi$ scattering up to high  
energies\footnote{
However, in the case of the Standard Model a single $\sigma$ suffices.}.
We will explore the quantum numbers of these resonances.

We now consider the scattering of the Goldstone bosons in the
Littlest Higgs ($SU(5)/SO(5)$) model in some detail, and show the
role a $\sigma$ resonance could play in unitarizing the theory.
Here we discuss the $\sigma$'s in the theory that does not
incorporate the techni-$\rho$s; it is possible to extend this
analysis to theories that model these states.

It is useful to decompose the $\pi\pi$ scattering amplitude into
representations of the unbroken symmetry group.   In the
Littlest Higgs model the $\pi\pi\rightarrow \pi\pi$ scattering can be
decomposed into various $SO(5)$ channels:
\begin{eqnarray}
\mathbf{14}\otimes\mathbf{14} \rightarrow
\mathbf{1} \oplus\mathbf{10}\oplus \mathbf{14} \oplus \cdots \, .
\end{eqnarray}
Each of the partial wave amplitudes has a different scale of
unitarity violation. It is the fluctuation in
the singlet channel (the direction of the vev)
that has the lowest scale of unitarity violation.  This is because
the other channels have smaller Clebsch-Gordon coefficients in the
decomposition.  A single, broad scalar should be expected
for the lightest $\sigma$ resonance, unitarizing this singlet channel.
Other representations are important for
staving off unitarity violation in the other channels.

We can find the relevant $\sigma$ model by considering a
linear $\sigma$ model for the breaking of $SU(5) \rightarrow SO(5)$.
To accomplish this breaking, a symmetric tensor of $SU(5)$
acquires a vev.   The symmetric tensor decomposes under $SO(5)$
as:
\begin{eqnarray}
30 \rightarrow  \mathbf{14}^- \oplus \mathbf{14}^+ \oplus
\mathbf{1}^+ \oplus \mathbf{1}^-.
\end{eqnarray}
We have introduced a ``charge conjugation'' $\mathbb{Z}_2$ symmetry
under which the $\rho$'s are even, but the $a_{1}$ is odd.  The
$\mathbf{1}^+$ is the state responsible for unitarizing the most  
strongly
coupled singlet channel.  The $\mathbf{14}^-$ are the pions of the
Little Higgs theory. As $N_f$ grows large, we believe that
the $\sigma$ becomes light and is responsible for unitarity violation.
This is very closely related the restoration of chiral symmetry at
large $N_f$ where one still has confinement but no chiral symmetry  
breaking.
A possible theory for the linear sigma model that displays this  
behavior is:
\begin{eqnarray}
\LL_\eff \sim |\partial \Phi |^2  - V(\Phi),
\hspace{0.5in}
V(\Phi) =  \lambda ( N_{f_c}- N_f ) ( |\Phi|^2 -  f^2)^2.
\end{eqnarray}
As $N_f$ approaches $N_{f_c}$ the $\sigma$ resonance becomes
light, after $N_{f_c}$ chiral symmetry is restored.
The dynamics is a continuous in $N_f$, and as $N_f$ approaches
the critical number of flavors to restore chiral symmetry
the $\sigma$ resonances become light and degenerate with the
$\pi$, filling out an entire chiral multiplet.  Therefore, in
the large $N_{f}$ limit, the scale of unitarity violation seems
closely tied to the presence of additional strongly coupled, broad
scalar resonances rather than new vector mesons.  New vector resonances
could cut-off gauge quadratic divergences.
If they are mandated to be light to unitarize the theory, then quadratic
divergences may be cut-off well beneath the NDA scale.  On the other hand,
if the $\sigma$ is unitarizing the theory, the vector resonances can
be heavy, and the divergences will be cut-off closer to the NDA scale.

\subsection{Unitarity Moral}

Regardless of whether we place our faith in the NDA
analysis or, alternately, the partial wave unitarity analysis,
it is clear that new modes are expected to appear beneath $4\pi f$.

In the previous sections we illustrated the effect of
the introduction of $\rho$-type resonances, showing that
they can soften the cut-off dependence of the theory, thus making
previously incalculable quantities calculable. In this section
we analyzed the role of these $\rho$'s in unitarizing $\pi-\pi$
scattering amplitude, as compared to that of $\sigma$-type states.

\begin{figure}
\begin{center}
\epsfig{file=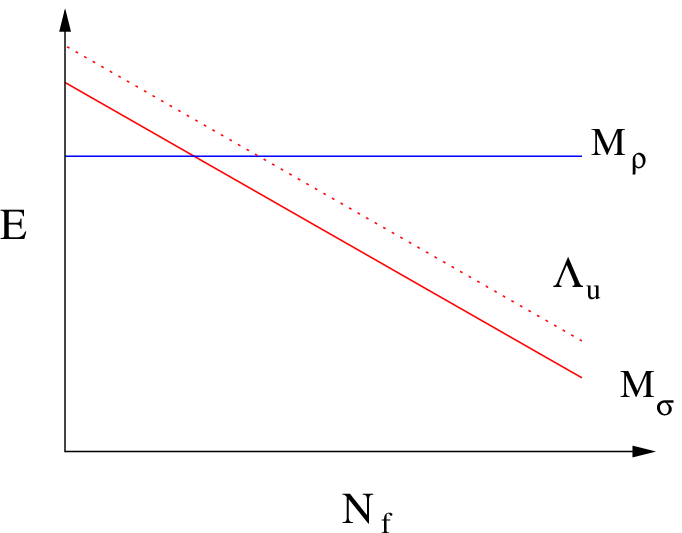, width=3.0in}
\hspace{0.1in}
\epsfig{file=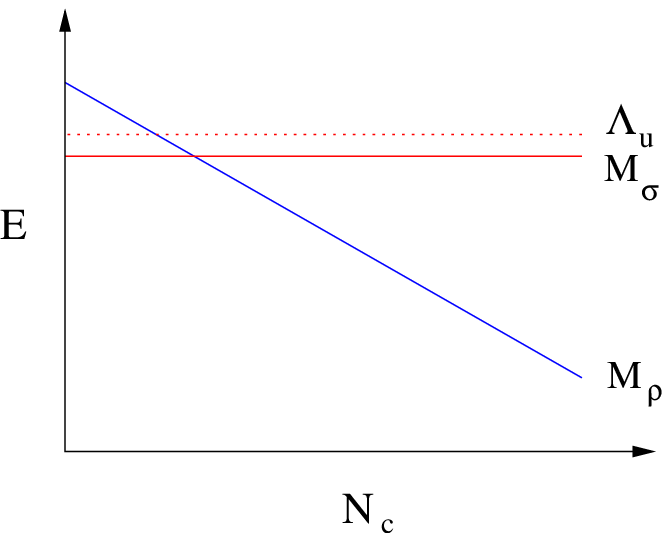, width=3.0in}
\caption{
\label{Fig: NewStates}
The left figure illustrates how the mass of the lightest $\sigma$
and $\rho$ and scale of unitarity violation in $\pi\pi$ scattering scale
with $N_f$.  The $\sigma$ resonances become light and unitarize the
scattering in a Higgs-like manner.
The right figure show how they vary with $N_c$.
The $\rho$ resonances become light (and simultaneously weakly
interacting) in the large $N_c$ limit
and can stave off unitarity violation as in Randall-Sundrum models
or Higgsless models.
}
\end{center}
\end{figure}

Our results are well illustrated by Fig~\ref{Fig: NewStates}.
It is clear that, as suggested by the intuition,
the scale of unitarization is strongly correlated
with the mass of the $\sigma$ fields.
A partial unitarization can be achieved with the introduction
of a tower of vector resonances.   If these resonances are present, the Goldstone boson 
scattering may remain well-behaved up to higher energies, even without the introduction
of a $\sigma$ resonance.   For the vector resonances to be important for unitarizing
scattering, they should light; however, from discussion about little Higgs models, this
possibility seems phenomenologically disastrous: they contribute to a light Higgs and a
large $S$ parameter.  Thus we expect broad $\sigma$ resonances for these models
to unitarize the scattering if Little Higgs models are to be phenomenologically viable.
 So while $\sigma$ resonances might be crucial for understanding unitarity, they are
 essentially invisible and do not affect the low energy phenomenology.
 
\section{Conclusions} 
\setcounter{equation}{0}
\renewcommand{\theequation}{\thesection.\arabic{equation}}
\label{Sec: Conclusion}

Using effective field theory techniques, we have studied vector resonances, 
moderately light relative to the scale of strong dynamics.  This approach 
allows the exploration of the vector limit, a point of enhanced symmetry.
Georgi's vector limit corresponds to a theory that is local in its theory 
space description.  
In the vector limit, the lightest $\rho$ regulates
the leading cut-off sensitive operators in the chiral Lagrangian.   In QCD, 
this corresponds to a softening of the divergence in the $\pi^{\pm}-\pi^{0}$ 
mass difference.  

It is not clear that Georgi's vector limit is in any way fundamental, and 
whether we expect it to hold in a generic (non-QCD) theory of strong 
coupling.  However, if it does apply, then it can have important 
phenomenological consequences.   We considered these implications 
by extrapolating this approach to the structure of techni-$\rho$s in Little Higgs 
theories.  By including techni-$\rho$s in these theories and assuming 
the analogue of the vector limit, we were able to discuss previously 
ultraviolet sensitive, phenomenologically relevant quantities.
For example, by using large $N_c$ arguments, we argued that the mass 
of the Higgs boson in the Littlest Higgs theory is roughly the mass
of $m_W$ and decreases as the techni-$\rho$s become light.  It should 
be noted that there are large radiative corrections from the top quark that
we did not estimate.

Finally, we briefly explored unitarity violation in Little Higgs models.
We argued that the scale of unitarity break-down likely points to 
broad $\sigma$-like
resonances.  If, instead of $\sigma$-like fields, this scale pointed to 
the presence of techni-$\rho$'s, then gauge quadratic divergences would be
cut off at this scale.  The result would be a too-light Higgs boson.  

We close with a few directions for further work.  
In principle, the techni-$\rho$ vector resonances can have masses 
similar to those of the additional gauge bosons of Little Higgs models.  This
could change the collider phenomenology predictions, and conceivably
modify precision electroweak predictions.  Using the formalism 
introduced here, it should be possible to pursue this question further.
Also, there is a UV completion for the Littlest Higgs that uses a
strongly coupled supersymmetric $SO(7)$ gauge theory
\cite{UVCompletion}.  
Using the ideas in this note detailed predictions of the
semi-perturbative regime could be analyzed.

\begin{figure}
\begin{center}
\epsfig{file=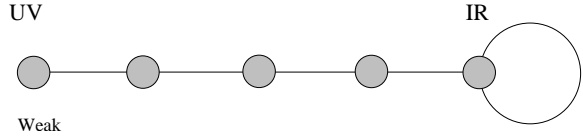, width=3.0in}
\caption{
\label{Fig: UVIR}
A schematic depiction of how the techni-$\rho$ resonances turn into 
deconstructed space maps into a holographic extra dimension.   The weakly gauged 
group is on the left.   The additional sites represent strongly gauged 
techni-$\rho$s.  In the IR, chiral symmetry is broken
by a non-linear sigma model field.
}
\end{center}
\end{figure}

These ideas of modeling the the techni-$\rho$ resonances are closely
related to deconstructing AdS \cite{AdS,DCAdS}.  As the number of sites grows large,
theory space reconstructs an extra-dimension as in Fig \ref{Fig: UVIR}.  It would be
interesting to explore this structure to see if deconstructing AdS
leads to some insight into the generalized vector limit.

\section*{Acknowledgments}

We would like to thank S. Chang, N. Arkani-Hamed, E. Katz, 
M. Luty, and M. Peskin for useful discussions.  
We would also like to acknowledge the Aspen Center for Physics where 
this work began.

\end{document}